\documentclass[aps,prd,superscriptaddress,amsmath,amsfont,amssymb,preprint,nofootinbib]{revtex4}
\pdfoutput=1
\usepackage{graphicx}

\usepackage[T1]{fontenc} 

\usepackage{epsfig,ulem,color,slashed}
\allowdisplaybreaks  

\usepackage[usenames,dvipsnames]{xcolor}
     \definecolor{hgreen}{rgb}{0,.3,0}
     \definecolor{hred}{rgb}{.3,0,0}
     \definecolor{hblue}{rgb}{0,0,.3}
     \definecolor{LightGray}{gray}{0.95}
\usepackage[
            colorlinks=true,
            linkcolor=hblue,
            citecolor=hgreen,
            filecolor=hblue,
            urlcolor=hred]{hyperref}

\newcommand{\lrpartial}{\negthickspace\stackrel{\leftrightarrow}{\partial}\negthickspace{}}

\newcommand{\beq}{\begin{equation} }
\newcommand{\eeq}{\end{equation}} 
\newcommand{\bi}{\begin{itemize} }
\newcommand{\ei}{\end{itemize} }
\newcommand{\C}{{\cal C}}
\newcommand{\Q}{{\cal Q}}
\newcommand{\op}{{\cal O}}

\definecolor{Red}{rgb}{1.,0.,0.}
\definecolor{Grn}{rgb}{0.,0.6,0.}
\definecolor{Blu}{rgb}{0.,0.,1.}
\definecolor{Pur}{rgb}{1.,0.,1.}

\begin{document} 

\title{\boldmath Effective Field Theory for Dark Matter Direct
  Detection up to Dimension Seven}

\def\TUD{Fakult\"at f\"ur Physik, TU Dortmund, D-44221 Dortmund, Germany}
\def\Cinci{Physics Department, University of Cincinnati, Cincinnati OH 45221, USA}

\author{\textbf{Joachim Brod}}
\email{joachim.brod@tu-dortmund.de}
\affiliation{\TUD}
\affiliation{\Cinci}

\author{\textbf{Aaron Gootjes-Dreesbach}}
\email{aaron.gootjes-dreesbach@tu-dortmund.de}
\affiliation{\TUD}

\author{\textbf{Michele Tammaro}}
\email{tammarme@mail.uc.edu}
\affiliation{\Cinci}

\author{\textbf{Jure Zupan}}
\email{zupanje@ucmail.uc.edu}
\affiliation{\Cinci}

\begin{abstract}
We present the full basis of effective operators relevant for dark
matter direct detection, up to and including operators of mass
dimension seven. We treat the cases where dark matter is either a
Dirac fermion, a Majorana fermion, a complex scalar, or a real scalar,
allowing for dark matter to furnish a general representation of the
electroweak gauge group. We describe the algorithmic procedure used to
obtain the minimal set of effective operators and provide the
tree-level matching conditions onto the effective theory valid below
the electroweak scale.
\end{abstract}

\preprint{DO-TH 17/21}

\maketitle

\tableofcontents

\flushbottom

\section{Introduction}
\label{sec:introduction}
In order to compare results of different Dark Matter (DM) direct
detection experiments one needs a theoretical framework in which to
interpret the data. Since the typical momentum exchanges between DM
and nuclei are small, $q_{\rm max}\lesssim 200$ MeV, the interactions
are well described within Effective Field Theory (EFT) without
recourse to detailed particle physics models \cite{Bishara:2016hek,
  Fan:2010gt, Fitzpatrick:2012ix, Fitzpatrick:2012ib,
  Anand:2013yka,DelNobile:2013sia, Barello:2014uda,
  Hill:2014yxa,Catena:2014uqa, Kopp:2009qt, Hill:2013hoa, Hill:2011be,
  Kurylov:2003ra, Pospelov:2000bq, Bagnasco:1993st, Cirigliano:2012pq,
  Hoferichter:2015ipa, Hoferichter:2016nvd, Bishara:2017pfq,
  Menendez:2012tm, Klos:2013rwa, Baudis:2013bba, Vietze:2014vsa}. The
interactions are ordered according to the dimensionality of the
operators, giving an effective interaction Lagrangian
\begin{equation}\label{eq:lightDM:Lnf5}
{\cal L}_\chi=\sum_{a,d}
\hat \C_{a}^{(d)} {\cal Q}_a^{(d)}\,, 
\qquad {\rm where}\quad 
\hat \C_{a}^{(d)}=\frac{\C_{a}^{(d)}}{\Lambda^{d-4}}\,.
\end{equation}
Here, $\C_a^{(d)}$ are dimensionless Wilson coefficients, $\Lambda$ is
the NP scale, generically of the order of the mediator mass, and the sum runs over the operator labels, $a$,
and the mass dimension of the operators, $d$. The EFT description is
valid as long as the mediators are heavier than the momenta exchanges,
$q_{\rm max}\ll \Lambda$.

The UV theories of DM interactions with the SM generically give a
nonzero contribution to at least one of the operators with $d\leq
7$. This set of operators already covers, for instance, all the
different chiral structures one can form out of two DM and two SM
fermion fields. Since the operators with low mass dimensions typically
give the dominant contributions to the scattering rates, it suffices
for most purposes to truncate the sum in \eqref{eq:lightDM:Lnf5} at
$d=7$. Indeed, the commonly used DM EFTs include operators up to mass
dimension seven~\cite{Goodman:2010qn, Bishara:2017nnn}. However, the
sets used in the literature are not complete -- they do not contain
the full basis of independent dimension-seven operators. In the
present manuscript we rectify this situation.

We construct a complete EFT basis for operators up to and including
dimension seven, coupling fermionic and scalar DM to the visible
sector. We provide two sets of operators: a basis valid at low
energies, $\mu\sim 2$ GeV, as well as a basis in the unbroken
electroweak phase, valid at or above the electroweak scale $\mu =
\mu_{\rm EW}\sim m_Z$. There are a number of operators at dimension
seven that were not considered before. We show that the complete basis
can be chosen in such a way that these additional operators contain
derivatives acting only on the DM currents. In the case of fermionic
DM such operators are likely to be only a subleading correction to the
operators already considered previously. The situation is
qualitatively different for scalar DM. Here, the operators with
derivatives acting on DM fields enter already at mass dimension six
and can form the leading contribution (for instance, if DM is a
pseudo-Nambu-Goldstone boson). Throughout this paper we restrict our
consideration to operators bilinear in the DM fields. To obtain a
basis that is closed under renormalization group (RG) evolution, one
would have to include operators with a different number of DM fields.

The formulation of the EFT at scale $\mu$ depends on which degrees of
freedom are relevant at that particular scale. As a starting point we
formulate in Section~\ref{subsec:low:EFT} the EFT for fermionic DM
that is valid at $\mu\sim 2$ GeV and describes the couplings of DM to
the $u,d,s$ quarks, leptons, gluons, and photons. In
Section~\ref{subsec:nonrel} we then perform a nonperturbative
matching, at $\mu\sim 2$ GeV, to an EFT that describes the couplings
of DM to nonrelativistic protons and neutrons.
Section~\ref{sec:RGrunning} contains the details about the
renormalization group running from the electroweak scale down to
$\mu\sim$ 2 GeV. In Section~\ref{sec:high:scale} we provide the
necessary ingredients to relate to the EFT above the electroweak
symmetry-breaking scale. In Section~\ref{sec:EFT:EWK} we first present
the complete basis of dimension-seven operators above the electroweak
scale, while in Section~\ref{sec:matching} we perform the matching at
the electroweak scale onto the five-flavor EFT by integrating out the
Higgs, $W$, and $Z$ bosons and the top quark. In
Section~\ref{sec:scalarDM} we extend the above formalism to the case
of scalar DM. Section~\ref{sec:conclusions} contains our
conclusions. Finally, in Appendix~\ref{sec:basis} we spell out the
details on the construction of the operator bases, while in
Appendix~\ref{sec:NR:app} we collect the relevant results for the
nonrelativistic reduction of the DM and nucleon fields.

\section{Low-scale interactions of fermionic DM}
\label{sec:fermionicDM}
We start by considering Dirac fermion DM, with field $\chi$, and write
down the complete basis of dimension-five, dimension-six, and
dimension-seven operators at $\mu \sim 2\,$GeV, for interactions
between DM and quarks, gluons, photons, and leptons,
Eq.~\eqref{eq:lightDM:Lnf5}. When writing the basis we assume that
there is a conserved global dark $U(1)_D$ quantum number, which
forbids currents of the form $\bar \chi^c \Gamma \chi$, where $\chi^c$
is the charge-conjugated DM field, and $\Gamma$ denotes a generic
string of Dirac matrices. The obtained basis is valid at all scales
below the electroweak scale, $\mu\lesssim M_W$, if matching
corrections at the charm- and bottom-quark thresholds and RG running
are taken into account, as detailed in
Section~\ref{sec:RGrunning}. The nonperturbative matching onto an EFT
coupling DM to nonrelativistic nucleons is presented in
Section~\ref{subsec:nonrel}. We also comment on the modifications
needed when considering Majorana DM.

\subsection{Fermionic DM coupling to quarks, gluons and photons}
\label{subsec:low:EFT}
The basis for dimension-five and dimension-six operators coincides
with the one in Refs.~\cite{Bishara:2017nnn, Bishara:2017pfq}. The two
dimension-five operators are the magnetic and electric dipole
operators,
\begin{align}
\label{eq:Q5:12}
\mathcal{Q}^{(5)}_1 &=  \frac{e}{8\pi^2} (\bar{\chi}\sigma^{\mu\nu}\chi) F_{\mu\nu}\,, &\quad
\mathcal{Q}^{(5)}_2 &=  \frac{e}{8\pi^2} (\bar{\chi}\sigma^{\mu\nu}i\gamma_5\chi) F_{\mu\nu}\,,
\end{align}
where $F^{\mu\nu} \equiv \partial_\mu A_\nu - \partial_\nu A_\mu$ is
the electromagnetic field-strength tensor, and
$\sigma^{\mu\nu} = \tfrac{i}{2} [\gamma^\mu, \gamma^\nu]$. The four
dimension-six operators are formed from products of vector and
axial-vector currents,
\begin{align}
\label{eq:Q6:1f}
\mathcal{Q}^{(6)}_{1,f} &=  (\bar{\chi}\gamma^\mu\chi)\, (\bar{f} \gamma_\mu f)\,, &\quad
\mathcal{Q}^{(6)}_{2,f} &=  (\bar{\chi}\gamma^\mu\gamma_5\chi)\, (\bar{f} \gamma_\mu f)\,,
 \\
\mathcal{Q}^{(6)}_{3,f} &=  (\bar{\chi}\gamma^\mu\chi)\, (\bar{f} \gamma_\mu\gamma_5 f)\,, &\quad
\mathcal{Q}^{(6)}_{4,f} &=  (\bar{\chi}\gamma^\mu\gamma_5\chi)\, (\bar{f} \gamma_\mu\gamma_5 f)\,. 
\end{align}
The subscript $f$ runs over the quark and lepton fields with masses
$m_f < \mu$, so that $f \in
\{u,d,s,e,\mu,\tau,\nu_e,\nu_\mu,\nu_\tau\}$. (For SM neutrinos only
the left-handed combination appears.)  The above basis is easily
extended to all scales up to the electroweak breaking scales. For
scales $m_c\lesssim \mu \lesssim m_b$, in addition the charm quark is
a propagating degree of freedom, so the operators with $f=c$ need to
be included, and for $m_b\lesssim \mu\lesssim \mu_{\rm EW}$ also the
operators involving the bottom-quark field, $f=b$.

At dimension seven, there are four operators involving products of
gluon field strength tensors
$G^{a\mu\nu} \equiv \partial_\mu G_\nu^a - \partial_\nu G_\mu^a + g_s
f^{abc} G_\mu^b G_\nu^c$,
\begin{align}
\label{eq:Q7:1:2}
\mathcal{Q}^{(7)}_1 &=  \frac{\alpha_s}{12\pi} (\bar{\chi}\chi)\, G^{a\mu\nu} G^a_{\mu\nu}\,, &\quad
\mathcal{Q}^{(7)}_2 &=  \frac{\alpha_s}{12\pi} (\bar{\chi}i\gamma_5\chi)\, G^{a\mu\nu} G^a_{\mu\nu}\,, 
\\
\label{eq:Q7:3:4}
\mathcal{Q}^{(7)}_3 &=  \frac{\alpha_s}{8\pi} (\bar{\chi}\chi)\, G^{a\mu\nu} \tilde{G}^a_{\mu\nu}\,, &\quad
\mathcal{Q}^{(7)}_4 &=  \frac{\alpha_s}{8\pi} (\bar{\chi}i\gamma_5\chi)\, G^{a\mu\nu} \tilde{G}^a_{\mu\nu}\,,
\end{align}
where $\tilde{G}^a_{\mu\nu} = \frac{1}{2}\varepsilon_{\mu\nu\rho\eta}
G_a^{\rho\eta}$. There are also six operators involving a scalar or
tensor fermion current. As they naturally arise from operators which,
above the electroweak scale, involve a Higgs field, we include an
explicit factor of $m_f$ and count them as dimension seven (see
Section~\ref{sec:high:scale}):
\begin{align}
\label{eq:Q7:5:6}
\mathcal{Q}^{(7)}_{5,f} &= m_f (\bar{\chi}\chi)\, (\bar{f} f)\,,
&\quad \mathcal{Q}^{(7)}_{6,f} &= m_f (\bar{\chi}i\gamma_5\chi)\,
(\bar{f} f)\,, \\
\label{eq:Q7:7:8}
\mathcal{Q}^{(7)}_{7,f} &=  m_f (\bar{\chi}\chi)\, (\bar{f} i\gamma_5 f)\,, &\quad
\mathcal{Q}^{(7)}_{8,f} &=  m_f (\bar{\chi}i \gamma_5\chi)\, (\bar{f} i \gamma_5 f)\,, 
\\
\label{eq:Q7:9:10}
\mathcal{Q}^{(7)}_{9,f} &=  m_f (\bar{\chi}\sigma^{\mu\nu}\chi)\, (\bar{f} \sigma_{\mu\nu} f)\,, &\quad
\mathcal{Q}^{(7)}_{10,f} &=  m_f (\bar{\chi}\sigma^{\mu\nu}i\gamma_5\chi)\, (\bar{f} \sigma_{\mu\nu} f)\,.
\end{align}
The operators in Eqs.~\eqref{eq:Q7:1:2}-\eqref{eq:Q7:9:10} were
already used in Ref.~\cite{Bishara:2017pfq}. The full basis of
dimension-seven operators includes, in addition, the Rayleigh
operators coupling DM to two photon field strength tensors,
\begin{align}
\label{eq:Q7:11}
\mathcal{Q}^{(7)}_{11} &= \frac{\alpha}{12\pi}(\bar{\chi}\chi)\,
F^{\mu\nu} F_{\mu\nu}\,, &\quad \mathcal{Q}^{(7)}_{12} &=
\frac{\alpha}{12\pi}(\bar{\chi}i\gamma_5\chi)\, F^{\mu\nu}
F_{\mu\nu}\,, \\
\label{eq:Q7:13}
\mathcal{Q}^{(7)}_{13} &=  \frac{\alpha}{8\pi}(\bar{\chi}\chi)\, F^{\mu\nu} \tilde{F}_{\mu\nu}\,, &\quad
\mathcal{Q}^{(7)}_{14} &=  \frac{\alpha}{8\pi}(\bar{\chi}i\gamma_5\chi)\, F^{\mu\nu} \tilde{F}_{\mu\nu}\,,
\end{align}
and four four-fermion operators with derivatives acting on the DM
currents,\footnote{There are four more independent structures closely
  related to the operators in Eqs.~\eqref{eq:4Fd7:3}
  and~\eqref{eq:4Fd7:4}, namely,
  $m_f (\bar{\chi} \gamma_\nu \chi)\, \partial_\mu (\bar{f}
  \sigma^{\mu\nu} f)$,
  $m_f (\bar{\chi} \gamma_\nu \chi)\, \partial_\mu (\bar{f}
  \sigma^{\mu\nu} i\gamma_5 f)$,
  $m_f (\bar{\chi} \gamma_\nu \gamma_5 \chi)\, \partial_\mu (\bar{f}
  \sigma^{\mu\nu} f)$, and
  $m_f (\bar{\chi} \gamma_\nu \gamma_5 \chi)\, \partial_\mu (\bar{f}
  \sigma^{\mu\nu} i\gamma_5 f)$.  These involve a chirality-flipping
  fermion current. Therefore, following our counting rules, they get
  multiplied by the corresponding fermion masses, rendering them
  dimension eight.}
\begin{align}
\mathcal{Q}^{(7)}_{15,f}
 &=  \partial_\mu(\bar{\chi}\sigma^{\mu\nu} \chi)\,
     (\bar{f} \gamma_\nu  f)\,,
 &\quad \mathcal{Q}^{(7)}_{16,f}
 &=  \partial_\mu(\bar{\chi}\sigma^{\mu\nu}i\gamma_5 \chi)\,
     (\bar{f} \gamma_\nu  f)\,, \label{eq:4Fd7:3}
\\
\mathcal{Q}^{(7)}_{17,f}
 &=  \partial_\mu(\bar{\chi}\sigma^{\mu\nu} \chi)\,
     (\bar{f} \gamma_\nu\gamma_5   f)\,,
 &\quad \mathcal{Q}^{(7)}_{18,f}
 &= \partial_\mu(\bar{\chi}\sigma^{\mu\nu}i\gamma_5 \chi)\,
    (\bar{f} \gamma_\nu\gamma_5 f)\,. \label{eq:4Fd7:4} 
\end{align}

This completes the EFT basis of dimension-seven operators for
fermionic DM that couple DM bilinears to the fields in the visible
sector. We have checked in two different ways that this set of
operators forms a complete basis: first, using an algorithmic
procedure that explicitly implements the equations of motion and
various algebraic identities, outlined in Appendix~\ref{sec:basis},
and second, with a procedure based on the conformal Hilbert
series~\cite{Lehman:2015via, Henning:2015daa, Henning:2015alf}.

Note that for Majorana DM the operators $\Q_{1,2}^{(5)}$,
$\Q_{1,f}^{(6)}$, $\Q_{3,f}^{(6)}$, $\Q_{9,f}^{(7)}$,
$\Q_{10,f}^{(7)}$, and $\Q_{15,f}^{(7)}, \ldots, \Q_{18,f}^{(7)}$
vanish, while the definitions of all the other operators coventionally
include an additional factor of $1/2$ (see also Appendix~A of
Ref.~\cite{Bishara:2017pfq}).

\subsection{RG running}
\label{sec:RGrunning}
If the initial conditions for Wilson coefficients are set in the
four-flavor EFT, $m_c\lesssim \mu\lesssim m_b$, or in the five-flavor
EFT, $m_b\lesssim \mu \lesssim \mu_{\rm EW}$, the Wilson coefficients
$\C_i^{(d)}$ need to be evolved down to $\mu\sim 2$ GeV. The RG
running for operators $\Q_1^{(5)}, \ldots, \Q_{10,f}^{(7)}$ in
Eqs.~\eqref{eq:Q5:12}-\eqref{eq:Q7:9:10} can be read off from
Ref.~\cite{Hill:2014yxa}. The effect of weak mixing below the weak
scale, potentially important for dimension-six operators,
Eq.~\eqref{eq:Q6:1f}, was presented in Ref.~\cite{Brod:2018ust},
including the leading logarithmic QCD corrections. Note that the
additional dimension-seven operators with derivatives on DM fields,
$\Q_{15,f}^{(7)}, \ldots, \Q_{18,f}^{(7)}$ in
  Eqs.~\eqref{eq:4Fd7:3} and~\eqref{eq:4Fd7:4}, have vanishing
one-loop anomalous dimension, and there are no matching corrections at
the quark thresholds.

The CP-even Rayleigh operators $\Q_{11}^{(7)}$, $\Q_{12}^{(7)}$ in
Eq.~\eqref{eq:Q7:11} mix into the scalar operators $\Q_{5,f}^{(7)},
\Q_{6,f}^{(7)}$ in Eq.~\eqref{eq:Q7:5:6} with anomalous dimension
\begin{equation}
  \gamma_{11;5,f} = \gamma_{12;6,f} = 8 Q_f^2 \bigg( \frac{\alpha}{4\pi} \bigg)^2 \,,
\end{equation}
where the running is determined by the RG equations
\begin{equation}
  \mu\frac{d}{d\mu} C_{5,f}^{(7)} (\mu) = C_{11}^{(7)} (\mu)
  \gamma_{11;5,f} \qquad \text{and} \qquad \mu\frac{d}{d\mu}
  C_{6,f}^{(7)} (\mu) = C_{12}^{(7)} (\mu) \gamma_{12;6,f} \,.
\end{equation}
The CP-odd Rayleigh operators $ \mathcal{Q}^{(7)}_{13}, \Q^{(7)}_{14}$
in Eq.~\eqref{eq:Q7:13} do not mix into the pseudo-scalar operators in
Eq.~\eqref{eq:Q7:7:8}, contrary to their QCD counterparts, due to the
absence of a QED anomaly. The RG evolution of the operators in
Eqs.~\eqref{eq:Q5:12}-\eqref{eq:4Fd7:4} is implemented in the {\tt
  DirectDM} code~\cite{Bishara:2017nnn}.

\subsection{Couplings to nonrelativistic nucleons and nuclear response}
\label{subsec:nonrel}
The scattering of DM on nuclei is dominated by DM scattering on single
nucleons. To leading order in the chiral expansion the scattering is
therefore described by the interactions of DM with nonrelativistic
nucleons and protons for which the EFT Lagrangian
is~\cite{Anand:2013yka, Bishara:2017nnn},
\begin{equation}\label{eq:LNR}
{\cal L}_{\rm NR}=\sum_{i,N} c_i^N(q^2) \op_i^N\,,
\end{equation}
with two momentum-independent nonrelativistic
operators,
\begin{align}
\label{eq:O1pO4p}
{\mathcal O}_1^N&= {1}_\chi {1}_N\,,
&{\mathcal O}_4^N&= \vec S_\chi \cdot \vec S_N \,,
\end{align}
and a set of momentum-dependent operators (displaying only the ones needed below),
\begin{align}
\label{eq:O5pO6p}
{\mathcal O}_5^N&= \vec S_\chi \cdot \Big(\vec v_\perp \times \frac{i\vec q}{m_N} \Big) \, {1}_N \,,
&{\mathcal O}_6^N&= \Big(\vec S_\chi \cdot \frac{\vec q}{m_N}\Big) \, \Big(\vec S_N \cdot \frac{\vec q}{m_N}\Big),
\\
\label{eq:O7pO8p}
{\mathcal O}_7^N&= {1}_\chi \, \big( \vec S_N \cdot \vec v_\perp \big)\,,
&{\mathcal O}_9^N&= \vec S_\chi \cdot \Big(\frac{i\vec q}{m_N} \times \vec S_N \Big)\,,
\\
{\mathcal O}_{11}^N&= - \Big(\vec S_\chi \cdot \frac{i\vec q}{m_N} \Big) \, {1}_N \,,
&{\mathcal O}_{14}^N&= -\Big(\vec S_\chi \cdot \frac{i\vec q}{m_N}  \Big) \, \Big(\vec S_N\cdot \vec v_\perp  \Big) \,,
\end{align}
where $N=p,n$. The momentum exchanges are defined as
\begin{equation}
\vec q = \vec k_2-\vec k_1=\vec p_1 -\vec p_2\,, \qquad \vec v_\perp=
\frac{\vec p_1+\vec p_2}{2{m_\chi}} - \frac{\vec k_1+\vec
  k_2}{2{m_N}}\,,
\end{equation}
where $p_{1(2)}$ is the momentum of incoming (outgoing) DM particle,
and similarly $k_{1(2)}$ for incoming (outgoing) nucleon.

The chirally leading nonrelativistic reduction of the
operators~\eqref{eq:Q5:12}-\eqref{eq:Q7:9:10} was already given in
Refs.~\cite{Bishara:2016hek,Bishara:2017pfq}. The nonrelativistic
reduction for the remaining dimension-seven operators with quark
currents is 
\begin{align}
\begin{split}
Q_{15,q}^{(7)} \to&  \, \frac{\vec q^{\,\,2}}{2 m_\chi} F_1^{q/N} \op_{1}^N +\frac{2 \vec q^{\,\,2}}{m_N}\big(F_1^{q/N}+F_2^{q/N}\big)\op_4^N
\\
& -2 m_N F_1^{q/N} \op_5^N-2 m_N\big(F_1^{q/N}+F_2^{q/N}\big)\op_6^N\,,
\end{split}
\\
Q_{16,q}^{(7)} \to&  -2 m_N F_1^{q/N} \op_{11}^N\,,
\\
Q_{17,q}^{(7)} \to&  -4 m_N F_A^{q/N} \op_{9}^N\,,
\\
Q_{18,q}^{(7)} \to&  \,4 m_N F_A^{q/N} \op_{14}^N\,.
\end{align}

The vector and axialvector form factors are defined through
\begin{align}
\label{vec:form:factor}
\langle N'|\bar q \gamma^\mu q|N\rangle&=\bar u_N'\Big[F_1^{q/N}(q^2)\gamma^\mu+\frac{i}{2m_N}F_2^{q/N}(q^2) \sigma^{\mu\nu}q_\nu\Big]u_N\,,
\\
\label{axial:form:factor}
\langle N'|\bar q \gamma^\mu \gamma_5 q|N\rangle&=\bar u_N'\Big[F_A^{q/N}(q^2)\gamma^\mu\gamma_5+\frac{1}{2m_N}F_{P'}^{q/N}(q^2) \gamma_5 q^\mu\Big]u_N\,,
\end{align}
where our notation suppresses the dependence of nucleon states on
their momenta, i.e., $\langle N'|\equiv\langle N(k_2)| $,
$| N\rangle\equiv | N(k_1)\rangle $, and
$\bar u_N'\equiv \bar u_N(k_2)$, $u_N\equiv u_N(k_1)$.

The contributions of the operators $Q_{15,q}^{(7)}, \ldots,
Q_{18,q}^{(7)}$ to the coefficients $c_i^N$ in the nonrelativistic
Lagrangian are thus 
\begin{align}
c_1^N& = F_1^{q/N} \frac{\vec q^{\,\,2}}{2 m_\chi} \C_{15,q}^{(7)} + \cdots \,,
\\
c_4^N& = \frac{2 \vec q^{\,\,2}}{m_N}\big(F_1^{q/N}+F_2^{q/N}\big) \C_{15,q}^{(7)} + \cdots \,,
\\
c_5^N& = -2 m_N F_1^{q/N} \C_{15,q}^{(7)} + \cdots \,,
\\
c_6^N& = -2 m_N\big(F_1^{q/N}+F_2^{q/N}\big) \C_{15,q}^{(7)} + \cdots \,,
\\
c_9^N& = -4 m_N F_A^{q/N} \C_{17,q}^{(7)} + \cdots \,,
\\
c_{11}^N& = -2 m_N F_1^{q/N} \C_{16,q}^{(7)} + \cdots \,,
\\
c_{14}^N& = 4 m_N F_A^{q/N} \C_{18,q}^{(7)} \,.
\end{align}
The ellipsis denotes the contribution of the operators in
Eqs.~\eqref{eq:Q5:12}-\eqref{eq:Q7:9:10}, with the results given in
Eqs.~(17)-(26) of Ref.~\cite{Bishara:2017nnn}. The coefficient
$c_{14}^N$ receives only contributions from dimension-seven
operators. In the above expressions we do not show the contributions
of the Rayleigh operators, since their hadronic matrix elements are
poorly known~\cite{Ovanesyan:2014fha}. The coefficient $c_1^N$, for
instance, receives a nonperturbative contribution from the CP-even
Rayleigh operator, $Q_{11}^{(7)}\sim (\chi\bar \chi)FF$ in
Eq.~\eqref{eq:Q7:11}. In the EFT with nonrelativistic nucleons one
would need to keep both this contribution to $\op_1^N$, as well as the
Rayleigh operator $Q_{11}^{(7)}$ itself, with appropriately modified
Wilson coefficient due to the matching.  The operators
$Q_{12}^{(7)}\sim(\chi\bar i \gamma_5 \chi)F F$,
$Q_{13}^{(7)}\sim(\chi\bar \chi)F\tilde F$, and
$Q_{14}^{(7)}\sim(\chi\bar i \gamma_5 \chi)F \tilde F$ similarly give
contributions in the nonperturbative matching to the non-relativistic
operators $\op_{11}^N \sim \vec S_\chi\cdot \vec q$, $\op_{10}^N\sim
\vec S_N\cdot \vec q$, and $\op_{6}^N \sim (\vec S_\chi\cdot \vec
q\,)(\vec S_N\cdot \vec q\,)$, respectively. The matching also
modifies the Wilson coefficients of these Rayleigh operators. At
present only NDA estimates for the effects of the matching are
available~\cite{Ovanesyan:2014fha, Appelquist:2015zfa, Weiner:2012cb,
  Frandsen:2012db}.

This completes the matching of the most general EFT Lagrangian for
fermionic DM coupling to SM through operators up to and including
dimension seven. The cross section for DM scattering on nuclei, due to
DM scattering on a single nucleon, is given by Eqs.~(40) and~(46) of
Ref.~\cite{Anand:2013yka}.  These expressions do receive large
corrections from the 2-nucleon interactions in the case of Rayleigh
operators. They are of the size $\sim Z Q_0/m_N$ relative to the
single-nucleon scattering, and can be dominant for large nuclei. Here
$Q_0\sim 1/(r_0 A^{1/3})\sim 160~{\rm MeV}/A^{1/3}$ is the momentum
scale corresponding to the size of the nucleus, see
Ref.~\cite{Ovanesyan:2014fha} for further details. For all the other
UV operators the two-nucleon contributions are subleading, see, e.g.,
Ref.~\cite{Bishara:2016hek}, and the above expressions can be readily
used to obtain the leading-order expressions in chiral counting for
the DM--nucleus scattering cross section.

\section{Interactions for fermionic DM above electroweak scale}
\label{sec:high:scale}

If the mediators are heavier than the electroweak symmetry-breaking
scale it is more convenient to write the operators in a way that is
manifestly invariant under the SM gauge group. This extends the SM-EFT
Lagrangian~\cite{Buchmuller:1985jz, Grzadkowski:2010es} by adding the
operators that include also DM fields.

Here, we allow DM to be part of a multiplet of the electroweak
$SU(2)$, with a possibly nonzero hypercharge, $Y_\chi$. We denote the
multiplet field by $\chi$ and its electromagnetically neutral
component\footnote{In Section~\ref{sec:fermionicDM} we did not display
  the superscript ``0'' in order to shorten the notation.} (the DM) by
$\chi^0$. We assume the DM interactions to be invariant under a global
dark $U(1)_D$ symmetry, with the DM multiplet charged under $U(1)_D$.

The size of the EFT operator basis depends on the dimensionality of DM
SU(2) multiplet, $d_\chi$.  One needs to distinguish three different
cases. For a DM multiplet with $d_\chi\geq 3$ the basis is the most
general one, and is given below. If DM is a doublet, $d_\chi=2$, one
can use the completeness relation for Pauli matrices,
\begin{equation}
\label{eq:completeness}
\sigma_{ij}^a \sigma_{kl}^a=2 \delta_{il}\delta_{kj}-\delta_{ij}\delta_{kl},
\end{equation}
to reduce the basis to a smaller set. Below we comment on which of the
operators should be dropped for the case of $d_\chi=2$ in order to
obtain a complete set of operators without redundant members. For
singlet DM, $d_\chi=1$, further reductions of the basis occur: because
the SU(2) generators in that representation are zero, several
operators vanish trivially.

\subsection{EFT basis above the electroweak scale}
\label{sec:EFT:EWK}
We start the construction of EFT for DM interactions above the
electroweak symmetry breaking scale by discussing the renormalizable
interactions, i.e., the kinetic term in the Lagrangian
\begin{equation}\label{eq:L:EW}
{\cal L}=i\bar \chi \slashed D\chi - m_\chi \bar \chi \chi.
\end{equation}
If DM is an electroweak singlet, $d_\chi=1$, $Y_\chi=0$, the covariant
derivative is simply a partial derivative. For DM in a nontrivial
representation of electroweak group the covariant derivative contains
the renormalizable interactions between DM and the $W^\pm$, $Z$ gauge
bosons. In the important special case of vanishing hypercharge of DM,
$Y_\chi=0$, the neutral component $\chi^0$ does not couple to the $Z$
boson at tree level. This suppresses the direct detection rates below
the current experimental bounds.

For such cases a phenomenologically interesting possibility is that
the leading interactions between DM and the visible sector, relevant
for direct detection experiments, are due to higher dimension
operators. We collect them in the effective Lagrangian
\begin{equation}\label{eq:Lagr:EW}
{\cal L}_{\rm EW}=\sum_{a,d} \frac{C_a^{(d)}}{\Lambda^{d-4}}
Q_a^{(d)} \,.
\end{equation}
Here, the label $a$ runs over different operators of dimension
$d$, the $C_a^{(d)}$ are the corresponding dimensionless Wilson
coefficients, and $\Lambda$ may be identified with the mass of the
mediator. 

The number of independent operators depends on whether DM is an
$SU(2)$ singlet, a doublet, or has $d_\chi\geq 3$.  For general
multiplet, $d_\chi \geq 3$, there are 8 dimension-five operators, 18
dimension-six operators, and 100 dimension-seven operators the couple
DM currents to the visible sector, not counting multiplicities due to
the quark and lepton flavors, and choosing a basis where all operators
are Hermitian. The basis is smaller for $d_\chi=2$ $(d_\chi=1)$, with
8 (4) dimension-five, 18 (12) dimension-six, and 92 (50)
dimension-seven operators. The complete basis for dimension-five and
dimension-six operators of DM interacting with the SM can be found in
Ref.~\cite{Bishara:2018vix}, and we repeat it here for completeness.
We present the complete basis of the dimension-seven operators here
for the first time.

\subsubsection{Dimension-five operator basis} 
The CP-conserving dimension-five operators are
\begin{align}
Q_{1}^{(5)} &= \frac{g_1}{8\pi^2} (\bar\chi\sigma^{\mu\nu}\chi)B_{\mu\nu}\,,
&Q_{2}^{(5)} &= \frac{g_2}{8\pi^2} (\bar\chi\sigma^{\mu\nu}\tilde\tau^a\chi)W_{\mu\nu}^a\,, \label{Q12}\\
Q_{3}^{(5)} &= (\bar\chi\chi)(H^\dagger H)\,,
&Q_{4}^{(5)} &= (\bar\chi\tilde\tau^a\chi)(H^\dagger \tau^a H)\,,\label{Q34}
\end{align}
while the CP-odd operators have an extra insertion of $\gamma_5$, 
\begin{align}
Q_{5}^{(5)} &= \frac{g_1}{8\pi^2} (\bar\chi\sigma^{\mu\nu}i\gamma_5\chi)B_{\mu\nu}\,,
&Q_{6}^{(5)} &= \frac{g_2}{8\pi^2} (\bar\chi\sigma^{\mu\nu}\tilde\tau^ai\gamma_5 \chi)W_{\mu\nu}^a\,,\label{Q56}\\
Q_{7}^{(5)} &= (\bar\chi i\gamma_5 \chi)(H^\dagger H)\,,
&Q_{8}^{(5)} &= (\bar\chi\tilde\tau^a i\gamma_5 \chi)(H^\dagger \tau^a H)\,.\label{Q78}
\end{align}
Here and below, $H$ is the SM Higgs doublet. All $SU(2)$ (and, below,
also color) indices are not displayed explicitly and are assumed to be
contracted within the brackets. Note that if $\chi$ is a $SU(2)$
singlet, the operators $Q_{2}^{(5)}$, $Q_{4}^{(5)}$, $Q_{6}^{(5)}$,
and $Q_{8}^{(5)}$ are absent.  In a perturbative UV theory the
operators $Q_{1,2}^{(5)}$ and $Q_{5,6}^{(5)}$ are generated at one
loop, while the operators $Q_{3,4}^{(5)}$ and $Q_{7,8}^{(5)}$ are
typically generated at tree level. This expectation is reflected in
our normalization of the operators.

\subsubsection{Dimension-six operator basis}

At dimension six there are many more operators. We do not consider
flavor violating operators, keeping our discussion minimal.  For each
SM fermion generation, $i=1,2,3$, there are then eight operators that
are products of DM currents and quark currents,
\begin{align}
Q_{1,i}^{(6)} &= (\bar\chi\gamma_\mu \tilde\tau^a\chi)(\bar Q_L^i \gamma^\mu \tau^a Q_L^i)\,, & Q_{5,i}^{(6)} &= (\bar\chi\gamma_\mu \gamma_5 \tilde\tau^a\chi)(\bar Q_L^i \gamma^\mu \tau^a Q_L^i)\,, \label{eq:dim6:Q15}\\
Q_{2,i}^{(6)} &= (\bar\chi\gamma_\mu \chi)(\bar Q_L^i \gamma^\mu Q_L^i)\,, & Q_{6,i}^{(6)} &= (\bar\chi\gamma_\mu \gamma_5 \chi)(\bar Q_L^i \gamma^\mu Q_L^i)\,,\\
Q_{3,i}^{(6)} &= (\bar\chi\gamma_\mu \chi)(\bar u_R^i \gamma^\mu u_R^i)\,, & Q_{7,i}^{(6)} &= (\bar\chi\gamma_\mu \gamma_5 \chi)(\bar u_R^i \gamma^\mu u_R^i)\,,\\
Q_{4,i}^{(6)} &= (\bar\chi\gamma_\mu \chi)(\bar d_R^i \gamma^\mu d_R^i)\,, & Q_{8,i}^{(6)} &= (\bar\chi\gamma_\mu \gamma_5 \chi)(\bar d_R^i \gamma^\mu d_R^i)\,. \label{eq:dim6:Q48}
\end{align}
Here $Q_L$ denotes the left-handed quark doublet, and $u_R$, $d_R$ the
right-handed up- and down-type quark singlets, respectively. The
corresponding operators involving lepton currents are
\begin{align}
Q_{9,i}^{(6)} &= (\bar\chi\gamma_\mu \tilde\tau^a\chi)(\bar L_L^i \gamma^\mu \tau^a L_L^i)\,, & Q_{12,i}^{(6)} &= (\bar\chi\gamma_\mu \gamma_5 \tilde\tau^a\chi)(\bar L_L^i \gamma^\mu \tau^a L_L^i)\,,\label{eq:dim6:Q9Q12}\\
Q_{10,i}^{(6)} &= (\bar\chi\gamma_\mu \chi)(\bar L_L^i \gamma^\mu L_L^i)\,, & Q_{13,i}^{(6)} &= (\bar\chi\gamma_\mu \gamma_5 \chi)(\bar L_L^i \gamma^\mu L_L^i)\,,\\
Q_{11,i}^{(6)} &= (\bar\chi\gamma_\mu \chi)(\bar \ell_R^i \gamma^\mu \ell_R^i)\,, & Q_{14,i}^{(6)} &= (\bar\chi\gamma_\mu \gamma_5 \chi)(\bar \ell_R^i \gamma^\mu \ell_R^i)\,,\label{eq:dim6:Q11Q14}
\end{align}
where $L_L$ denotes the left-handed lepton doublet, and $\ell_R$ the
right-handed down-type lepton singlet. Finally, there are four
dimension-six operators involving Higgs currents,
\begin{align}
Q_{15}^{(6)} &= (\bar\chi\gamma^\mu \tilde\tau^a\chi)(H^\dagger
i\stackrel{\leftrightarrow}{D^a}_\mu H)\,, & Q_{17}^{(6)} &=
(\bar\chi\gamma^\mu \gamma_5 \tilde\tau^a\chi)(H^\dagger
i\stackrel{\leftrightarrow}{D^a}_\mu H)\,,
\label{eq:dim6:Q15Q17}
\\ 
Q_{16}^{(6)} &= (\bar\chi\gamma^\mu \chi)(H^\dagger
i\stackrel{\leftrightarrow}{D}_\mu H)\,, & Q_{18}^{(6)} &=
(\bar\chi\gamma^\mu \gamma_5 \chi)(H^\dagger
i\stackrel{\leftrightarrow}{D}_\mu H)\,. 
\label{eq:dim6:Q16Q18} 
\end{align}
The Higgs currents are defined in terms of hermitian combinations of
the covariant derivatives, $\stackrel{\leftrightarrow}{D}_\mu
\,\equiv\, D_\mu - \stackrel{\leftarrow}{D}_\mu^\dagger$ and
$\stackrel{\leftrightarrow}{D^a}_\mu \,\equiv\, \tau^a D_\mu -
\stackrel{\leftarrow}{D}_\mu^\dagger \tau^a$. Additional operators
with covariant derivatives acting on the DM fields vanish via the DM
equations of motion, up to total derivatives.  As in the case of
dimension-five operators, the basis simplifies if DM is a $SU(2)$
singlet. In this case, the operators $Q_{1,i}^{(6)}$, $Q_{5,i}^{(6)}$,
$Q_{9,i}^{(6)}$, $Q_{12,i}^{(6)}$, $Q_{15}^{(6)}$, and $Q_{17}^{(6)}$
vanish and should be dropped from the basis.

\subsubsection{Dimension-seven operator basis}
We group the dimension-seven operators into several classes, depending
on how many derivatives they contain and to which SM fields they
couple. Our convention for the covariant derivative is $D_\mu \psi_f=
(\partial_\mu + i g_s T^a G^a_\mu - i g_2 \tilde \tau^a W^a_\mu + i
g_1 Y_f B_\mu/2)\psi_f $, where $T^a$ is the generator of $SU(3)_c$,
$\tilde \tau^a$ the generator of $SU(2)_L$, both in the representation
furnished by $\psi_f$ (for a doublet we denote it by $\tau^a$), while
$Y_f$ is the hypercharge of $\psi_f$.  The dual field-strength tensor
is defined as $\tilde{G}^a_{\mu\nu} =
\frac{1}{2}\varepsilon_{\mu\nu\rho\eta} G_a^{\rho\eta}$.

\paragraph{Gauge-Gauge operators.} For $d_\chi \geq 3$ there are 22
dimension-seven operators that couple DM to two gauge field-strength
tensors:\begin{align} Q^{(7)}_{1V} &=
  \frac{\alpha_{s}}{12\pi}(\bar{\chi}\chi) G^a_{\mu\nu}G^{a,\mu\nu}\,,
  &\quad Q^{(7)}_{2V} &=
  \frac{\alpha_{s}}{12\pi}(\bar{\chi}i\gamma_5\chi)
                        G^a_{\mu\nu}G^{a,\mu\nu}\,, \label{eq:Q12V7}\\
          Q^{(7)}_{3V} &=\frac{\alpha_{s}}{8\pi} (\bar{\chi}\chi)
                         G^a_{\mu\nu}\tilde{G}^{a,\mu\nu}\,, &\quad
                                                               Q^{(7)}_{4V}
                      &=\frac{\alpha_{s}}{8\pi}
                        (\bar{\chi}i\gamma_5\chi)
                        G^a_{\mu\nu}\tilde{G}^{a,\mu\nu}\,, \label{eq:Q34V7}\\
          Q^{(7)}_{5V} &= \frac{\alpha_1}{12\pi}(\bar{\chi}\chi)
                         B_{\mu\nu}B^{\mu\nu}\,, &\quad Q^{(7)}_{6V}
                      &=
                        \frac{\alpha_1}{12\pi}(\bar{\chi}i\gamma_5\chi)
                        B_{\mu\nu}B^{\mu\nu}\,, \label{eq:Q56V7}\\
          Q^{(7)}_{7V} &= \frac{\alpha_1}{8\pi}(\bar{\chi}\chi)
                         B_{\mu\nu}\tilde{B}^{\mu\nu}, &\quad
                                                         Q^{(7)}_{8V}
                      &=
                        \frac{\alpha_1}{8\pi}(\bar{\chi}i\gamma_5\chi)
                        B_{\mu\nu}\tilde{B}^{\mu\nu}\,, \\
          Q^{(7)}_{9V} &= \frac{\alpha_{12}}{12\pi}(\bar{\chi}\tilde
                         \tau^a\chi) W^a_{\mu\nu}B^{\mu\nu}\,, &\quad
                                                                 Q^{(7)}_{10V}
                      &=
                        \frac{\alpha_{12}}{12\pi}(\bar{\chi}i\gamma_5\tilde
                        \tau^a\chi)
                        W^a_{\mu\nu}B^{\mu\nu}\,, \\
          Q^{(7)}_{11V} &= \frac{\alpha_{12}}{8\pi}(\bar{\chi}\tilde
                          \tau^a\chi)
                          W^a_{\mu\nu}\tilde{B}^{\mu\nu}\,, &\quad
                                                              Q^{(7)}_{12V}
                      &=
                        \frac{\alpha_{12}}{8\pi}(\bar{\chi}i\gamma_5\tilde
                        \tau^a\chi)
                        W^a_{\mu\nu}\tilde{B}^{\mu\nu}\,, \\
\label{eq:Q1314V7}
Q^{(7)}_{13V} &=
\frac{\alpha_{12}}{12\pi}(\bar{\chi}\sigma^{\mu\nu}\tilde \tau^a\chi)
W^a_{\mu\sigma}B_\nu^{\,\,\,\sigma}\,, &\quad Q^{(7)}_{14V} &=
\frac{\alpha_{12}}{12\pi}(\bar{\chi}\sigma^{\mu\nu}i\gamma_5\tilde
\tau^a\chi) W^a_{\mu\sigma}B_\nu^{\,\,\,\sigma}\,, 
\\ 
Q^{(7)}_{15V} &=
\frac{\alpha_{2}}{12\pi}(\bar{\chi}\chi) W^a_{\mu\nu}W^{a,\mu\nu}\,,
&\quad Q^{(7)}_{16V} &=
\frac{\alpha_{2}}{12\pi}(\bar{\chi}i\gamma_5\chi)
W^a_{\mu\nu}W^{a,\mu\nu}\,, 
\\ 
Q^{(7)}_{17V} &=
\frac{\alpha_{2}}{8\pi}(\bar{\chi}\chi)
W^a_{\mu\nu}\tilde{W}^{a,\mu\nu}\,, &\quad Q^{(7)}_{18V} &=
\frac{\alpha_{2}}{8\pi}(\bar{\chi}i\gamma_5\chi)
W^a_{\mu\nu}\tilde{W}^{a,\mu\nu}\,, 
\\ 
Q^{(7)}_{19V} &=
\frac{\alpha_{2}}{8\pi}(\bar{\chi}\sigma^{\mu\nu}\tilde \tau^a\chi)
W^b_{\mu\sigma}W^{c,\sigma}_{\nu} \epsilon^{abc} ,&\quad Q_{20V}^{(7)} &=
\frac{\alpha_{2}}{8\pi}(\bar{\chi}\sigma^{\mu\nu}i\gamma_5 \tilde
\tau^a\chi) W^b_{\mu\sigma}W^{c,\sigma}_{\nu} \epsilon^{abc}\,,
\label{eq:Q1920V7}
\\
\label{eq:Q2122V7}
Q^{(7)}_{21V} &=
\frac{\alpha_{2}}{12\pi}(\bar{\chi} \tilde \tau^{\{ab\}} \chi) W^a_{\mu\nu}W^{b,\mu\nu}\,,
&\quad Q^{(7)}_{22V} &= \frac{\alpha_{2}}{12\pi}(\bar{\chi}i\gamma_5 \tilde \tau^{\{ab\}} \chi) W^a_{\mu\nu}W^{b,\mu\nu}\,,
\end{align}
where
$\tilde \tau^{\{ab\}}=\tilde \tau^{a} \tilde \tau^{b}+\tilde \tau^{b}
\tilde \tau^{a}$. Among those, the eleven operators $\Q_{kV}^{(7)}$
with $k=1,4,5,8,9,12,13,15,18,19,21$ are CP even, while the remaining
eleven are CP odd. In the definitions of the operators we included
loop factors with appropriate gauge couplings,
\begin{equation}\label{eq:gauge:factors}
\alpha_1=\frac{g_1^2}{4\pi}, \qquad \alpha_2=\frac{g_2^2}{4\pi}\,, \qquad \alpha_{12}=\frac{g_1 g_2}{4\pi},\qquad \alpha_s=\frac{g_3^2}{4\pi}\,,
\end{equation}
since these operators arise in perturbative theories at one loop or higher. 

For $d_\chi<3$ not all of the above operators are linearly
independent. For DM that is an electroweak doublet, $d_\chi=2$, the
operators $\Q_{21V}^{(7)}, \Q_{22V}^{(7)}$ should be dropped from the
basis, since they are equivalent to $\Q_{15V}^{(7)},
\Q_{16V}^{(7)}$. Similarly, for DM that is an electroweak singlet,
$d_\chi=1$, the operators $\Q_{9V}^{(7)}, \ldots, Q_{14V}^{(7)}$ and
$\Q_{19V}^{(7)}, \ldots, \Q_{22V}^{(7)}$ should be dropped from the
basis.

\paragraph{Gauge-Higgs operators.} For $d_\chi \geq 3$ there are 12
operators that couple DM currents to a set of mixed Higgs and
gauge-boson operators,
\begin{align}
\label{eq:Q71V'2V'}
Q^{(7)}_{1V'} &= \frac{g_1}{8\pi^2}(\bar{\chi}\sigma^{\mu\nu}\chi)
B_{\mu\nu}\; H^\dagger H\,, 
&\ Q^{(7)}_{2V'} &= \frac{g_1}{8\pi^2}
(\bar{\chi}\sigma^{\mu\nu} i\gamma_5\chi) B_{\mu\nu}\; H^\dagger H\,,
\\
 Q^{(7)}_{3V'} &= \frac{g_1}{8\pi^2}(\bar{\chi}\sigma^{\mu\nu}\tilde
\tau^a\chi) B_{\mu\nu}\; H^\dagger \tau^a H\,, &\quad Q^{(7)}_{4V'} &=
\frac{g_1}{8\pi^2}(\bar{\chi}\sigma^{\mu\nu} i\gamma_5\tilde
\tau^a\chi) B_{\mu\nu}\; H^\dagger \tau^a H\,, \\ Q^{(7)}_{5V'} &=
\frac{g_2}{8\pi^2}(\bar{\chi}\sigma^{\mu\nu}\chi) W^a_{\mu\nu}\;
H^\dagger\tau^a H\,, &\quad Q^{(7)}_{6V'} &= \frac{g_2}{8\pi^2}
(\bar{\chi}\sigma^{\mu\nu} i\gamma_5\chi) W^a_{\mu\nu}\;
H^\dagger\tau^a H\,, \\ Q^{(7)}_{7V'} &=
\frac{g_2}{8\pi^2}(\bar{\chi}\sigma^{\mu\nu}\tilde \tau^a\chi)
W^a_{\mu\nu}\; H^\dagger H\,, &\quad Q^{(7)}_{8V'} &=
\frac{g_2}{8\pi^2}(\bar{\chi}\sigma^{\mu\nu} i\gamma_5\tilde
\tau^a\chi) W^a_{\mu\nu}\; H^\dagger H\,, \\
\label{eq:Q79V'10V'}
Q^{(7)}_{9V'} &=  \frac{g_2}{8\pi^2}(\bar{\chi}\sigma^{\mu\nu}\tilde \tau^a\chi) W^b_{\mu\nu}\;  H^\dagger\tau^c H \epsilon^{abc}\,, &
Q^{(7)}_{10V'} &=  \frac{g_2}{8\pi^2}(\bar{\chi}\sigma^{\mu\nu} i\gamma_5\tilde \tau^a\chi) W^b_{\mu\nu}\;  H^\dagger\tau^c H \epsilon^{abc}\,,
\\
\label{eq:Q711V'12V'}
Q^{(7)}_{11V'} &=  \frac{g_2}{8\pi^2}(\bar{\chi} \sigma^{\mu\nu} \tilde \tau^{\{ab\}} \chi) W^a_{\mu\nu}\;  H^\dagger\tau^b H \,, &
Q^{(7)}_{12V'} &=  \frac{g_2}{8\pi^2}(\bar{\chi} \sigma^{\mu\nu} i\gamma_5 \tilde \tau^{\{ab\}} \chi) W^a_{\mu\nu}\;  H^\dagger\tau^b H \,.
\end{align}
The definitions of the operators include an appropriate gauge coupling
and a loop factor, because they are expected to arise at one loop when
matching from a UV theory. For DM that is part of an electroweak
doublet, $d_\chi=2$, the operators $Q^{(7)}_{11V'}$ and
$Q^{(7)}_{12V'}$ should be dropped from the basis, while for singlet
DM, $d_\chi=1$, in addition the operators $Q^{(7)}_{3V',4V'}$, and
$Q^{(7)}_{7V',\ldots, 10V'}$ should be dropped.

\paragraph{Four-Higgs operators.} For $d_\chi\geq 3$ there are six
dimension-seven operators that involve only DM and Higgs fields,
\begin{align}
\label{eq:Q7:1H:2H}
Q^{(7)}_{1H} &=  (\bar{\chi}\chi) \; H^\dagger H \; H^\dagger H\,, &\quad
Q^{(7)}_{2H} &=  (\bar{\chi} i\gamma_5 \chi) \; H^\dagger H \; H^\dagger H\,, \\
\label{eq:Q7:3H:4H}
Q^{(7)}_{3H} &=  (\bar{\chi}\tilde \tau^a\chi) \; H^\dagger \tau^a H \; H^\dagger H\,, &\quad
Q^{(7)}_{4H} &=  (\bar{\chi} i\gamma_5 \tilde \tau^a\chi) \; H^\dagger \tau^a H \; H^\dagger H\,,
\\
\label{eq:Q7:5H:6H}
Q^{(7)}_{5H} &=  (\bar{\chi} \tilde \tau^{\{ab\}} \chi) \; (H^\dagger \tau^a H) \; (H^\dagger \tau^b H)\,, &\quad
Q^{(7)}_{6H} &=  (\bar{\chi} i\gamma_5 \tilde \tau^{\{ab\}} \tilde \chi) \; (H^\dagger \tau^a H) \; (H^\dagger \tau^b H)\,.
\end{align}
The operators $Q^{(7)}_{1H}$, $Q^{(7)}_{3H}$, and $Q^{(7)}_{5H}$ are
CP even, while $Q^{(7)}_{2H}$, $Q^{(7)}_{4H}$, and $Q^{(7)}_{6H}$ are
CP odd. For DM that is an electroweak singlet, $d_\chi=1$, the
operators $Q^{(7)}_{3H}, \ldots, Q^{(7)}_{6H}$ are redundant and
should be dropped, while for DM that is a doublet, $d_\chi=2$, only
the operators $Q^{(7)}_{5H}$ and $Q^{(7)}_{6H}$ should be dropped from
the basis.

\paragraph{Two-Higgs operators.} There are ten operators coupling a
$d_\chi\geq3$ DM multiplet to Higgs bilinears, given by
\begin{align}
\label{eq:Q7:1H':2H'}
Q^{(7)}_{1H'} &= (\bar{\chi} \chi) D_\mu H^\dagger D^\mu H\,, &\quad
Q^{(7)}_{2H'} &= (\bar{\chi} i\gamma_5 \chi) D_\mu H^\dagger D^\mu
H\,, \\ Q^{(7)}_{3H'} &= (\bar{\chi} \tilde \tau^a \chi) D_\mu
H^\dagger \tau^a D^\mu H\,, &\quad Q^{(7)}_{4H'} &= (\bar{\chi}
i\gamma_5\tilde \tau^a \chi) D_\mu H^\dagger \tau^a D^\mu H\,,
\\ Q^{(7)}_{5H'} &= i(\bar{\chi} \sigma^{\mu\nu} \chi) D_\mu H^\dagger
D_\nu H\,, &\quad Q^{(7)}_{6H'} &= i(\bar{\chi} \sigma^{\mu\nu}
i\gamma_5 \chi) D_\mu H^\dagger D_\nu H\,, \label{eq:Q7:5H':6H'} \\
\label{eq:Q7:7H':8H'}
Q^{(7)}_{7H'} &=  i(\bar{\chi} \sigma^{\mu\nu} \tilde \tau^a \chi)  D_\mu H^\dagger \tau^a D_\nu H\,, 
&
Q^{(7)}_{8H'} &=  i(\bar{\chi} \sigma^{\mu\nu} i\gamma_5 \tilde \tau^a \chi)  D_\mu H^\dagger \tau^a D_\nu H\,,
\\ \label{eq:Q7:9H':10H'} 
Q^{(7)}_{9H'} &= i(\bar{\chi} \sigma^{\mu\nu} \tilde \tau^{\{ab\}} \chi) (D_\mu \tau^a H^\dagger) (D_\nu \tau^b H)\,, 
&Q^{(7)}_{10H'} &= i(\bar{\chi} \sigma^{\mu\nu} \tilde \tau^{\{ab\}} i\gamma_5 \chi) (D_\mu \tau^a H^\dagger) (D_\nu \tau^b H)\,.
\end{align}
For DM that is an electroweak doublet, $d_\chi=2$, the operators
$Q^{(7)}_{9H'}$ and $Q^{(7)}_{10H'}$ are redundant and should be
dropped from the basis. Similarly, for electroweak singlet DM,
$d_\chi=1$, the operators $Q^{(7)}_{3H'}$, $Q^{(7)}_{4H'}$,
$Q^{(7)}_{7H',\ldots,10H'}$ should be dropped.

\paragraph{Yukawa-like operators.} The operators with
chirality-flipping currents on both the DM and the SM side are
\begin{align}
\label{eq:Q71Y}
Q^{(7)}_{1Y,i} &= (\bar{\chi}\chi) (\bar{Q}_L^i u_R^i \tilde H) \,, &
Q^{(7)}_{2Y,i} &= (\bar{\chi}i\gamma_5\chi) (\bar{Q}_L^i u_R^i \tilde
H)\,, \\ Q^{(7)}_{3Y,i} &= (\bar{\chi}\tilde \tau^a\chi) (\bar{Q}_L^i
u_R^i \tau^a \tilde H) \,, & Q^{(7)}_{4Y,i} &=
(\bar{\chi}i\gamma_5\tilde \tau^a\chi) (\bar{Q}_L^i u_R^i \tau^a
\tilde H)\,, \\ Q^{(7)}_{5Y,i} &= (\bar{\chi}\sigma_{\mu\nu}\chi)
(\bar{Q}_L^i \sigma^{\mu\nu} u_R^i \tilde H)\,, & Q^{(7)}_{6Y,i} &=
(\bar{\chi}\sigma_{\mu\nu}\tilde \tau^a \chi) (\bar{Q}_L^i
\sigma^{\mu\nu} \tau^a u_R^i \tilde H )\,, \label{eq:Q75Y} \\ Q^{(7)}_{7Y,i} &=
(\bar{\chi}\chi) (\bar{Q}_L^i d_R^i H) \,, & Q^{(7)}_{8Y,i} &=
(\bar{\chi}i\gamma_5\chi) (\bar{Q}_L^i d_R^i H)\,, \\ Q^{(7)}_{9Y,i}
&= (\bar{\chi}\tilde \tau^a\chi) (\bar{Q}_L^i d_R^i \tau^aH) \,, &
Q^{(7)}_{10Y,i} &= (\bar{\chi}i\gamma_5\tilde \tau^a\chi) (\bar{Q}_L^i
d_R^i \tau^a H)\,, \\ Q^{(7)}_{11Y,i} &=
(\bar{\chi}\sigma_{\mu\nu}\chi) (\bar{Q}_L^i \sigma^{\mu\nu} d_R^i
H)\,, & Q^{(7)}_{12Y,i} &= (\bar{\chi}\sigma_{\mu\nu}\tilde \tau^a
\chi) (\bar{Q}_L^i \sigma^{\mu\nu} \tau^a d_R^i H )\,,
\label{eq:Q711Y} \\ Q^{(7)}_{13Y,i} &= (\bar{\chi}\chi) (\bar{L}_L^i \ell_R^i H)\,,
&Q^{(7)}_{14Y,i} &= (\bar{\chi}i\gamma_5\chi) (\bar{L}_L^i \ell_R^i
H)\,, \\ Q^{(7)}_{15Y,i} &= (\bar{\chi} \tilde \tau^a \chi)
(\bar{L}_L^i \ell_R^i \tau^a H ) \,, &Q^{(7)}_{16Y,i} &=
(\bar{\chi}i\gamma_5 \tilde \tau^a \chi) (\bar{L}_L^i \ell_R^i \tau^a
H)\,, \\
\label{eq:Q717Y}
Q^{(7)}_{17Y,i} &=  (\bar{\chi}\sigma_{\mu\nu}\chi)  (\bar{L}_L^i \sigma^{\mu\nu}  \ell_R^i  H) \,, 
\qquad &Q^{(7)}_{18Y,i} &=  (\bar{\chi}\sigma_{\mu\nu} \tilde \tau^a \chi) (\bar{L}_L^i \sigma^{\mu\nu}  \ell_R^i  \tau^a H)\,,
\end{align}
where $\tilde H=i \sigma_2 H^*$ is the charge-conjugated Higgs
field. The above operators are not Hermitian so that they enter the
effective Lagrangian, Eq.~\eqref{eq:Lagr:EW}, with complex
coefficients. Their Hermitian conjugates have to be included,
modifying \eqref{eq:Lagr:EW} to
\begin{equation}\label{eq:Lagr:EWmod}
{\cal L}_{\rm EW}=\sum_{a,d} \frac{C_a^{(d)}}{\Lambda^{d-4}} Q_a^{(d)}
(+{\rm h.c.})\,,
\end{equation}
where the ``h.c.'' is present in the sum only for the operators
$Q^{(7)}_{1Y,i}, \ldots, Q^{(7)}_{18Y,i}$.  Alternatively, one could
work with a set of 36 Hermitian operators with real coefficients, by
defining $Q^{(7)}_{i+}=Q^{(7)}_{i }+\big(Q^{(7)}_{i }\big)^\dagger$,
$Q^{(7)}_{i-}=i Q^{(7)}_{i }-i \big(Q^{(7)}_{i }\big)^\dagger$, for
each of the operators in Eqs.~\eqref{eq:Q71Y}-\eqref{eq:Q717Y}. The
above set of operators $Q^{(7)}_{kY,i}$ is part of a complete basis
for both $d_\chi\geq 3$ and for $d_\chi=2$, while if DM is an
electroweak singlet, $d_\chi=1$, the operators with
$k=3,4,6,9,10,12,15,16,18$, should be dropped from the basis.

\paragraph{Four-Fermion operators.} Finally, we list the tensor
operators that involve four fermion fields and one derivative, which
for $d_\chi\geq3$ and $d_\chi=2$ are given by
\begin{align}
\label{eq:Q7:1F2F}
Q^{(7)}_{1F,i} &=  \partial_\mu(\bar{\chi}\sigma^{\mu\nu}\tilde \tau^a\chi)  (\bar{Q}_L^i \tau^a\gamma_\nu  Q_L^i )\,, &\quad
Q^{(7)}_{2F,i} &=  \partial_\mu(\bar{\chi}\sigma^{\mu\nu}i\gamma_5\tilde \tau^a\chi)  (\bar{Q}_L^i \tau^a\gamma_\nu  Q_L^i )\,,
\\
Q^{(7)}_{3F,i} &=  \partial_\mu(\bar{\chi}\sigma^{\mu\nu}\chi)  (\bar{Q}_L^i \gamma_\nu  Q_L^i )\,, &\quad
Q^{(7)}_{4F,i} &=  \partial_\mu(\bar{\chi}\sigma^{\mu\nu}i\gamma_5\chi)  (\bar{Q}_L^i \gamma_\nu  Q_L^i )\,, 
\\
Q^{(7)}_{5F,i} &=  \partial_\mu(\bar{\chi}\sigma^{\mu\nu}\chi)  (\bar{u}_R^i \gamma_\nu  u_R^i )\,, &\quad
Q^{(7)}_{6F,i} &=  \partial_\mu(\bar{\chi}\sigma^{\mu\nu}i\gamma_5\chi)  (\bar{u}_R^i \gamma_\nu  u_R^i )\,, 
\\
Q^{(7)}_{7F,i} &=  \partial_\mu(\bar{\chi}\sigma^{\mu\nu}\chi)  (\bar{d}_R^i \gamma_\nu  d_R^i )\,, &\quad
Q^{(7)}_{8F,i} &=  \partial_\mu(\bar{\chi}\sigma^{\mu\nu}i\gamma_5\chi)  (\bar{d}_R^i \gamma_\nu  d_R^i )\,, 
\\
Q^{(7)}_{9F,i} &=  \partial_\mu(\bar{\chi}\sigma^{\mu\nu}\tilde \tau^a\chi)  (\bar{L}_L^i \tau^a\gamma_\nu  L_L^i )\,, &\quad
Q^{(7)}_{10F,i} &=  \partial_\mu(\bar{\chi}\sigma^{\mu\nu}i\gamma_5\tilde \tau^a\chi)  (\bar{L}_L^i \tau^a\gamma_\nu  L _L^i)\,, 
\\
Q^{(7)}_{11F,i} &=  \partial_\mu(\bar{\chi}\sigma^{\mu\nu}\chi)  (\bar{L}_L^i \gamma_\nu  L_L^i )\,, &\quad
Q^{(7)}_{12F,i} &=  \partial_\mu(\bar{\chi}\sigma^{\mu\nu}i\gamma_5\chi)  (\bar{L}_L^i \gamma_\nu  L_L^i )\,, \\
\label{eq:Q7:13F14F}
Q^{(7)}_{13F,i} &=  \partial_\mu(\bar{\chi}\sigma^{\mu\nu}\chi)  (\bar{\ell}_R^i \gamma_\nu  \ell_R^i )\,,&\quad
Q^{(7)}_{14F,i} &=  \partial_\mu(\bar{\chi}\sigma^{\mu\nu}i\gamma_5\chi)  (\bar{\ell}_R^i \gamma_\nu  \ell_R^i ) \,.
\end{align}
The operators $Q^{(7)}_{1F,i}$, $Q^{(7)}_{2F,i}$ and $Q^{(7)}_{9F,i}$,
$Q^{(7)}_{10F,i}$ should be dropped from the basis for singlet
DM, $d_\chi=1$. 

This completes the list of dimension-seven operators coupling
Dirac-fermion DM currents to the visible sector. If DM is a Majorana
particle, several of these operators vanish identically. These are the
operators that involve a DM vector or tensor current, namely the
gauge-gauge operators $Q^{(7)}_{13V}$, $Q^{(7)}_{14V}$,
$Q^{(7)}_{19V}$, $Q^{(7)}_{20V}$, in Eqs.~\eqref{eq:Q1314V7},
\eqref{eq:Q1920V7}; all of the gauge-Higgs operators
$Q^{(7)}_{1V'},\ldots, Q^{(7)}_{12V'}$, in
Eqs.~\eqref{eq:Q71V'2V'}-\eqref{eq:Q711V'12V'}, the two-Higgs
operators $Q^{(7)}_{5H'},\ldots, Q^{(7)}_{10H'}$, in
Eqs.~\eqref{eq:Q7:5H':6H'}-\eqref{eq:Q7:9H':10H'}; the Yukawa-like
operators $Q^{(7)}_{5Y,i}, Q^{(7)}_{6Y,i}$,
$Q^{(7)}_{11Y,i}, Q^{(7)}_{12Y,i}$, $Q^{(7)}_{17Y,i}, Q^{(7)}_{18Y,i}$
in Eqs.~\eqref{eq:Q75Y},~\eqref{eq:Q711Y}, and~\eqref{eq:Q717Y}; and
all of the four-fermion operators
$Q^{(7)}_{1F,i}, \ldots, Q^{(7)}_{14F,i}$ in
Eqs.~\eqref{eq:Q7:1F2F}-\eqref{eq:Q7:13F14F}.

\subsection{Tree-level matching at the electroweak scale}
\label{sec:matching}

Next, we perform the tree-level matching of the EFT for DM
interactions above the electroweak scale onto the EFT valid below the
electroweak scale. In the matching at $\mu\sim \mu_{EW}$ we integrate
out the heavy states with masses of the order of the electroweak scale
-- the Higgs boson, the top quark, and the $W$ and $Z$ bosons --
resulting in the effective Lagrangian~\eqref{eq:lightDM:Lnf5} with
five active quark flavors.  The results for the gauge-invariant EFT
containing dimension-five and dimension-six operators can be found in
Ref.~\cite{Bishara:2018vix}. Here, we present the additional contributions
due to the presence of dimension-seven operators.

We use the conventions of Ref.~\cite{Denner:1991kt}. The gauge fields
in the broken phase are
\begin{equation}
W_\mu^\pm = \frac{1}{2} \big( W_\mu^1 \mp i W_\mu^2\big)\,, \qquad
\begin{pmatrix}Z_\mu\\A_\mu\end{pmatrix} = 
\begin{pmatrix}c_w&s_w\\-s_w&c_w\end{pmatrix}
\begin{pmatrix}W_\mu^3\\B_\mu\end{pmatrix},
\end{equation}
with the sine of the weak mixing angle $s_w \equiv
g_1/\sqrt{g_1^2+g_2^2}$. Accordingly, we choose the following explicit
form of the $SU(2)$ generators
\begin{equation}\begin{split} \label{eq:su2-gen-def}
\big(\tilde \tau^1 \pm i \tilde\tau^2 \big)_{kl} = \delta_{k,l\pm 1}
\sqrt{(I_\chi \mp l)(I_\chi \pm l + 1)} \, , \qquad \big( \tilde\tau^3
\big)_{kl} = l \delta_{k,l} \, ,
\end{split}\end{equation}
with $I_\chi = (d_\chi-1)/2$, and the indices $k,l$ running over the
values $-I_\chi,\ldots , I_\chi - 1, I_\chi$. The Higgs doublet field
is
\begin{equation}
H =
\begin{pmatrix}G^+\\\tfrac{1}{\sqrt{2}} \big( v + h + i
  G^0 \big)\end{pmatrix}, 
\end{equation}
where $h$ denotes the physical Higgs field, and $G^\pm$, $G^0$ the
pseudo-Goldstone fields, eaten by the longitudinal components of
$W^\pm$ and $Z$.  

Before proceeding we remark that both the DM mass, $m_\chi$, and the
DM field $\chi$ get shifted by the contributions of the four-Higgs
operators in Eqs.~\eqref{eq:Q7:1H:2H}-\eqref{eq:Q7:5H:6H} if all Higgs
fields are replaced by their vacuum expectation values. The kinetic
and mass terms in the dimension-four Lagrangian Eq.~\eqref{eq:L:EW}
thus become
\begin{equation}\label{eq:redef-mass-field}
{\cal L}_\chi^{(4)}|_{n_f=5}=i \bar \chi' \slashed \partial \chi'
-m_\chi' \bar \chi' \chi'\,,
\end{equation}
where the $\chi$ field now only denotes the DM field, i.e., the
neutral component of the DM multiplet, which in addition has been
transformed by a simple chiral rotation, $\chi'= \exp\big({\frac{i}{2}
  \gamma_5 \phi}\big)\chi$, where (see also
Ref.~\cite{Fedderke:2014wda})
\begin{equation}\label{eq:chiral-rotation-angle}
\tan \phi = \frac{C_7^{(5)} + \tfrac{Y_\chi}{4}C_8^{(5)} +
  \Big(C_{2H}^{(7)} + \frac{Y_\chi}{4} C_{4H}^{(7)} + \frac{Y_\chi^{2}}{8}
  C_{6H}^{(7)}\Big) \frac{v_\text{EW}^2}{2\Lambda^2} }{ 2 m_\chi
  \Lambda / v_\text{EW}^2 - C_3^{(5)} - \tfrac{Y_\chi}{4}C_4^{(5)} -
  \Big( C_{1H}^{(7)} + \frac{Y_\chi}{4} C_{3H}^{(7)} +
  \frac{Y_\chi^{2}}{8} C_{5H}^{(7)} \Big)
  \frac{v_\text{EW}^2}{2\Lambda^2}} \,,
\end{equation}  
while the new mass term is
\begin{equation}\label{eq:mchi:shift}
\begin{split}
m_\chi'  =  m_\chi \cos\phi & - \frac{v^2}{2\Lambda}
\bigg[\bigg(C_7^{(5)}+\frac{Y_\chi}{4} C_8^{(5)}\bigg) \sin \phi +
  \bigg(C_3^{(5)}+\frac{Y_\chi}{4} C_4^{(5)}\bigg) \cos \phi \bigg]
  \\
   & - \frac{v^4}{4\Lambda^3} \bigg[\bigg(
  C_{2H}^{(7)} + \frac{Y_\chi}{4} C_{4H}^{(7)} + \frac{Y_\chi^{2}}{8}
  C_{6H}^{(7)} \bigg) \sin \phi \\ 
  & \qquad \qquad + \bigg( C_{1H}^{(7)} +
  \frac{Y_\chi}{4} C_{3H}^{(7)} + \frac{Y_\chi^{2}}{8} C_{5H}^{(7)} \bigg)
  \cos \phi \bigg]\,.
\end{split}
\end{equation}
The corrections from dimension-five and dimension-six operators were
already derived in Ref.~\cite{Bishara:2018vix}.

The field redefinition also changes the operators $Q_1^{(5)}, \dots,
Q_8^{(5)}$ in \eqref{Q12}--\eqref{Q78} and the corresponding Wilson
coefficients, $C_i^{(5)}{}'=C_i^{(5)} \cos \phi + C_{i+4}^{(5)} \sin
\phi \,$, $C_{i+4}^{(5)}{}' = C_{i+4}^{(5)} \cos \phi - C_{i}^{(5)}
\sin \phi \,,$ for $i=1,\dots, 4$, while there is no change in the
dimension-six Wilson coefficients. The field redefinition changes all
of the dimension-seven operators in this section. The corresponding
Wilson coefficients of the operators with primed fields are given by
\begin{equation}
C_{2k-1,A(2k,A)}^{(7)\prime}  = C_{2k-1,A(2k,A)}^{(7)} \cos\phi \pm C_{2k,A(2k-1,A)}^{(7)} \sin\phi,
\end{equation}
for $A=V (V',H,H',F)$, in which case $k=1,\ldots,11 (6,3,5,7)$, and
also for $A=Y$, with $k=1,2,4,5,7,8$, while for the remaining
dimension-seven Yukawa-like operators one has
\begin{align}
C_{k,Y}^{(7)\prime} & = C_{k,Y}^{(7)} \exp(i\phi) \,, & k = 5,6,11,12,17,18\,.
\end{align}
From now on we will assume that the above field and mass redefinitions
have been performed, and drop the primes on the Wilson coefficients,
the $\chi$ fields, and the DM mass $m_\chi$.

Next we give the contributions from dimension-seven operators above
the electroweak scale, when matching onto the EFT below the
electroweak scale (the contributions from dimension-five and
dimension-six operators are given in Ref.~\cite{Bishara:2018vix}). The
gauge-Higgs operators, Eqs.~\eqref{eq:Q71V'2V'}-\eqref{eq:Q79V'10V'},
contribute to the dimension-five dipole operators after the Higgs
obtains its vacuum expectation value, leading to
\begin{align}
\hat \C^{(5)}_{1(2)} &= \frac{v_{\rm EW}^2}{2
  \Lambda^3}\left[C^{(7)}_{1(2)V'} +\frac{Y_\chi}{4} C^{(7)}_{3(4)V'}
  +\frac{1}{2} C^{(7)}_{5(6)V'} +\frac{Y_\chi}{2}
  C^{(7)}_{7(8)V'}\right]+\ldots\,.  
\end{align}
The ellipsis denotes the tree-level contributions from the
dimension-five UV operators, as well as the one-loop contributions
from renormalizable gauge interactions. The explicit expressions for
both of these contributions can be found in Ref.~\cite{Bishara:2018vix}.
Note that there are no tree-level contributions from the operators
$Q^{(7)}_{9V'}$ and $Q^{(7)}_{10V'}$.

The matching of dimension-seven electroweak
operators~\eqref{eq:Q12V7}, \eqref{eq:Q34V7} onto the gluonic
operators~\eqref{eq:Q7:1:2}, \eqref{eq:Q7:3:4} is given by
\begin{equation}
\hat \C^{(7)}_{k} = \frac{1}{\Lambda^3} \cdot C^{(7)}_{(k)V}+\ldots\,,
\qquad k=1,\ldots, 4\,.
\end{equation}
The ellipsis denotes the one-loop contribution of the dimension-five
scalar operators, given in Ref.~\cite{Bishara:2018vix}.

The tree-level matching of the operators
\eqref{eq:Q56V7}-\eqref{eq:Q2122V7} onto the Rayleigh operators
\eqref{eq:Q7:11}-\eqref{eq:Q7:13} is given by 
\begin{align}
\hat \C^{(7)}_{11+k} &= \frac{1}{\Lambda^3}\left[C^{(7)}_{(5+k)V}
  +\frac{Y_\chi}{2}  C^{(7)}_{(9+k)V} + C^{(7)}_{(15+k)V}+\big(\delta_{k0}+\delta_{k1}\big)\frac{Y_\chi^2}{2}C^{(7)}_{(21+k)V}\right],
\qquad k=0,\ldots, 3\,. 
\end{align}
Note that there are no tree-level contributions to the interactions of
the neutral component~$\chi^0$ from the operators $Q^{(7)}_{13V},
Q^{(7)}_{14V}$, $Q^{(7)}_{19V}$, and $Q^{(7)}_{20V}$.

\begin{figure}[t]\centering
	\includegraphics[scale=0.8]{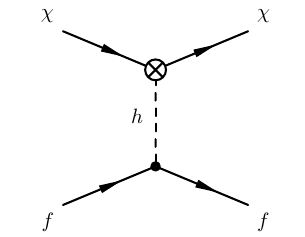}
	\caption{Matching contributions to dimension-seven effective
          operators for $\mu<v_{\rm EW}$ due to Higgs exchange.}
	\label{fig:d7-higgs-matching}
\end{figure}

The Yukawa-like operators, Eq.~\eqref{eq:Q71Y}-\eqref{eq:Q717Y}, match
onto the dimension-seven scalar-current four-fermion operators,
\eqref{eq:Q7:5:6}-\eqref{eq:Q7:9:10}, by replacing the Higgs with its
vacuum expectation value. In addition, the four-Higgs operators in
Eqs.~\eqref{eq:Q7:1H:2H}-\eqref{eq:Q7:5H:6H} contribute to the same
operators via tree-level Higgs exchange, see
Fig.~\ref{fig:d7-higgs-matching}. This yields the following
coefficients for up-type quarks ($u_1=u, u_2=c$)
\begin{align}
\begin{split}
\label{eq:C7:56ui}
\hat \C^{(7)}_{5(6),u_i} &= \frac{1}{\Lambda^3}\left[\frac{v_{\rm
      EW}}{\sqrt{2} m_{u_i}} \text{Re} \Big( C^{(7)}_{1(2)Y,i} - \frac{Y_\chi}{4} C^{(7)}_{3(4)Y,i}\Big)\right.
\\      
&\qquad\qquad\left.-\frac{v_{\rm
    EW}^2}{M_h^2}\Big(C^{(7)}_{1(2)H}+\frac{Y_\chi}{4} C^{(7)}_{3(4)H}+\frac{Y_\chi^2}{8} C^{(7)}_{5(6)H} \Big)\right]+\cdots\,,
\end{split}
\\
\hat \C^{(7)}_{7(8),u_i} & = \frac{1}{\Lambda^3}\frac{v_{\rm EW}}{
  \sqrt{2} m_{u_i}} \text{Im} \left( C^{(7)}_{1(2)Y,i} - \frac{Y_\chi}{4} C^{(7)}_{3(4)Y,i} \right)\,,\\
\hat \C^{(7)}_{9,u_i} &=\frac{1}{\Lambda^3} \frac{v_{\rm EW}}{
  \sqrt{2} m_{u_i}} \text{Re} \left( C^{(7)}_{5Y,i} - \frac{Y_\chi}{4} C^{(7)}_{6Y,i}  \right) \,,
\\
\hat \C^{(7)}_{10,u_i} &=\frac{1}{\Lambda^3} \frac{v_{\rm EW}}{
  \sqrt{2} m_{u_i}} \text{Im} \left( C^{(7)}_{5Y,i} - \frac{Y_\chi}{4} C^{(7)}_{6Y,i}  \right) \,,
\end{align}
where the ellipsis in Eq.~\eqref{eq:C7:56ui} denotes the contributions
from dimension-five operators, given in
Ref.~\cite{Bishara:2018vix}. For down-type quarks ($d_1=d, d_2=s, d_3=b$)
one has
\begin{align}
\begin{split}
\label{eq:C7:56di}
\hat \C^{(7)}_{5(6),d_i} &= \frac{1}{\Lambda^3}\left[\frac{v_{\rm
      EW}}{\sqrt{2} m_{d_i}} \text{Re} \Big( C^{(7)}_{7(8)Y,i} + \frac{Y_\chi}{4} C^{(7)}_{9(10)Y,i}\Big)\right.
      \\
&\qquad\qquad\left.-\frac{v_{\rm EW}^2}{M_h^2}\Big(C^{(7)}_{1(2)H}+\frac{Y_\chi}{4} C^{(7)}_{3(4)H}+\frac{Y_\chi^2}{8} C^{(7)}_{5(6)H} \Big)\right]+\cdots\,,
\end{split}
\\
\hat \C^{(7)}_{7(8),d_i} & = \frac{1}{\Lambda^3}\frac{v_{\rm EW}}{
  \sqrt{2} m_{d_i}} \text{Im} \left( C^{(7)}_{7(8)Y,i} + \frac{Y_\chi}{4} C^{(7)}_{9(10)Y,i} \right)\,,\\
\hat \C^{(7)}_{9,d_i} &=\frac{1}{\Lambda^3} \frac{v_{\rm EW}}{
  \sqrt{2} m_{d_i}} \text{Re} \left( C^{(7)}_{11Y,i} + \frac{Y_\chi}{4} C^{(7)}_{12Y,i}  \right) \,,
\\
\hat \C^{(7)}_{10,d_i} &=\frac{1}{\Lambda^3} \frac{v_{\rm EW}}{
  \sqrt{2} m_{d_i}} \text{Im} \left( C^{(7)}_{11Y,i} + \frac{Y_\chi}{4} C^{(7)}_{12Y,i}  \right) \,,
\end{align}
while for charged leptons ($\ell_1=e, \ell_2=\mu, \ell_3=\tau$)
\begin{align}
\begin{split}
\label{eq:C7:56li}
\hat \C^{(7)}_{5(6),\ell_i} &= \frac{1}{\Lambda^3}\left[\frac{v_{\rm
      EW}}{\sqrt{2} m_{\ell_i}} \text{Re} \Big( C^{(7)}_{13(14)Y,i} + \frac{Y_\chi}{4} C^{(7)}_{15(16)Y,i}\Big)\right.\\
&\qquad\qquad\left. -\frac{v_{\rm
      EW}^2}{M_h^2}\Big(C^{(7)}_{1(2)H}+\frac{Y_\chi}{4} C^{(7)}_{3(4)H}+\frac{Y_\chi^2}{8} C^{(7)}_{5(6)H}  \Big)\right]+\cdots\,,
\end{split}
\\
\hat \C^{(7)}_{7(8),\ell_i} & = \frac{1}{\Lambda^3}\frac{v_{\rm EW}}{
  \sqrt{2} m_{\ell_i}} \text{Im} \left( C^{(7)}_{13(14)Y,i} + \frac{Y_\chi}{4} C^{(7)}_{15(16)Y,i} \right)\,,\\
\hat \C^{(7)}_{9,\ell_i} &=\frac{1}{\Lambda^3} \frac{v_{\rm EW}}{
  \sqrt{2} m_{\ell_i}} \text{Re} \left( C^{(7)}_{17Y,i} + \frac{Y_\chi}{4} C^{(7)}_{18Y,i}  \right) \,,
\\
\hat \C^{(7)}_{10,\ell_i} &=\frac{1}{\Lambda^3} \frac{v_{\rm EW}}{
  \sqrt{2} m_{\ell_i}} \text{Im} \left( C^{(7)}_{17Y,i} + \frac{Y_\chi}{4} C^{(7)}_{18Y,i}  \right) \,.
\end{align}

The dimension-seven four-fermion UV operators in
Eq.~\eqref{eq:Q7:1F2F}-\eqref{eq:Q7:13F14F} match onto the
dimension-seven four-fermion operators in
Eq.~\eqref{eq:4Fd7:3}-\eqref{eq:4Fd7:4} after electroweak symmetry
breaking. The coefficients for up-type quarks are given by 
\begin{align}
\hat \C^{(7)}_{15(16),u_i} &= \frac{1}{2\Lambda^3} \left( -\frac{Y_\chi}{4} C^{(7)}_{1(2)F,i} + C^{(7)}_{3(4)F,i} + C^{(7)}_{5(6)F,i} \right)\,, \\
\hat \C^{(7)}_{17(18),u_i} &= \frac{1}{2 \Lambda^3} \left( \frac{Y_\chi}{4} C^{(7)}_{1(2)F,i} - C^{(7)}_{3(4)F,i} +C^{(7)}_{5(6)F,i} \right)\,,
\end{align}
while for down-type quarks one has 
\begin{align}
\hat \C^{(7)}_{15(16),d_i} &= \frac{1}{2\Lambda^3} \left( \frac{Y_\chi}{4} C^{(7)}_{1(2)F,i} + C^{(7)}_{3(4)F,i} + C^{(7)}_{7(8)F,i} \right)\,, \\
\hat \C^{(7)}_{17(18),d_i} &= \frac{1}{2 \Lambda^3} \left( -\frac{Y_\chi}{4} C^{(7)}_{1(2)F,i} - C^{(7)}_{3(4)F,i} + C^{(7)}_{7(8)F,i} \right)\,. 
\end{align}
The matching for charged leptons ($\ell_1=e, \ell_2=\mu, \ell_3=\tau$)
is given by 
\begin{align}
\hat \C^{(7)}_{15(16),\ell_i} &= \frac{1}{2\Lambda^3} \left(\frac{Y_\chi}{4} C^{(7)}_{9(10)F,i} + C^{(7)}_{11(12)F,i} + C^{(7)}_{13(14)F,i} \right)\,, \\
\hat \C^{(7)}_{17(18),e_i} &= \frac{1}{2\Lambda^3} \left( -\frac{Y_\chi}{4} C^{(7)}_{9(10)F,i} - C^{(7)}_{11(12)F,i} +C^{(7)}_{13(14)F,i} \right)\,,
\end{align}
and for neutrinos ($\nu_1=\nu_e, \nu_2=\nu_\mu, \nu_3=\nu_\tau$)
\begin{align}
\hat \C^{(7)}_{15(16),\nu_i} &= -\hat \C^{(7)}_{17(18),\nu_i}=\frac{1}{2 \Lambda^3} \left( - \frac{Y_\chi}{4} C^{(7)}_{9(10)F,i} + C^{(7)}_{11(12)F,i} \right)\,.
\end{align}
Note that the two-Higgs operators in
Eqs.~\eqref{eq:Q7:1H':2H'}-\eqref{eq:Q7:7H':8H'} do not contribute to
the matching at tree level.

\section{Scalar DM}
\label{sec:scalarDM}
Next we turn our attention to complex scalar DM, again allowing for DM
to be part of an electroweak multiplet (we comment on the changes
required for scalar DM below). There are only two renormalizable
couplings between scalar DM and the SM,
\begin{equation}
{\cal L}_{\rm EW}^\varphi \supset
- \frac{\lambda_{\varphi H}}{4} (\varphi^\dagger \varphi) (H^\dagger H)
- \frac{\lambda_{\varphi H}'}{4} (\varphi^\dagger\tilde \tau^a \varphi) (H^\dagger \tau^a H)\,.
\end{equation}
If these Higgs-portal couplings are absent or are very small, the
higher dimension operators can become important. This can happen, for
instance, if $\varphi$ is a pseudo-Goldstone boson in which case the
above two operators are forbidden by a shift symmetry.

The nonrenormalizable interactions of scalar DM with the visible
sector start at dimension six,
\begin{equation}\label{eq:Lagr:EW:scal}
{\cal L}_{\rm EW}^\varphi=\sum_{a,d} \frac{\C_a^{(d)}}{\Lambda^{d-4}} Q_a^{(d)}(+{\rm h.c.})\,.
\end{equation}
Note that, in order to simplify the notation, we denote the operators
and Wilson coefficients with the same symbols as we did for fermionic
DM in the previous two sections. The Hermitian conjugate in the sum is
present only when the operators are not Hermitian. This is the case
for the operators in Eqs.~\eqref{eq:Q18i:Q19i}-\eqref{eq:Q22i:Q23i},
whose Wilson coefficients can be complex.

\subsection{Scalar DM above the electroweak scale}
\label{sec:scal:basis:ew}
Above the electroweak scale there are 36 Hermitian operators at
dimension six that couple scalar DM in a $d_\chi \geq 3$ multiplet to
the currents in the visible sector. For $d_\chi=2$ this reduces to a
basis of 33 linearly independent operators, while for $d_\chi=1$ the
basis is further reduced to only 20 operators (not counting the flavor
structures of quark currents). The operators that couple DM currents
to the field strength tensors of the SM gauge fields are
\begin{align}
&Q_1^{(6)} =  \frac{\alpha_s}{12\pi}\left( \varphi^\dagger \varphi \right) G_{\mu\nu}^aG^{a,\mu\nu}\,,
 &\qquad  &Q_2^{(6)} =  \frac{\alpha_s}{8\pi}\left( \varphi^\dagger \varphi \right) G_{\mu\nu}^a\tilde{G}^{a,\mu\nu}\,,
 \\
&Q_3^{(6)} = \frac{\alpha_1}{12\pi}\left( \varphi^\dagger \varphi \right) B_{\mu\nu}B^{\mu\nu}\,, 
&
\qquad  &Q_4^{(6)} =  \frac{\alpha_1}{8\pi}\left( \varphi^\dagger \varphi \right) B_{\mu\nu}\tilde{B}^{\mu\nu}\,, \\
&Q_5^{(6)} =  \frac{\alpha_{12}}{12\pi}\left( \varphi^\dagger \tilde\tau^a \varphi \right) B_{\mu\nu}W^{a,\mu\nu} \,,
&\qquad  &Q_6^{(6)} =  \frac{\alpha_{12}}{8\pi}\left( \varphi^\dagger \tilde\tau^a \varphi \right) B_{\mu\nu}\tilde{W}^{a,\mu\nu}\,, \\
&\Q_7^{(6)} =  \frac{\alpha_2}{12\pi}\left( \varphi^\dagger \varphi \right) W_{\mu\nu}^aW^{a,\mu\nu}\,, 
&\qquad  &Q_8^{(6)} = \frac{\alpha_2}{8\pi}\left( \varphi^\dagger \varphi \right) W_{\mu\nu}^a\tilde{W}^{a,\mu\nu}\,, \\
&\Q_9^{(6)} = \frac{\alpha_2}{12\pi}\left( \varphi^\dagger \tilde \tau^{\{ab\}} \varphi \right) W_{\mu\nu}^aW^{b,\mu\nu} \,,
&\qquad  &\Q_{10}^{(6)} = \frac{\alpha_2}{8\pi}\left( \varphi^\dagger \tilde \tau^{\{ab\}} \varphi \right) W_{\mu\nu}^a\tilde{W}^{b,\mu\nu} \,,
\end{align}
where we included a loop factor with appropriate gauge couplings, see
Eq.~\eqref{eq:gauge:factors}. For DM that is an electroweak doublet,
$d_\chi=2$, the operators $\Q_9^{(6)}$ and $\Q_{10}^{(6)}$ should be
dropped from the basis as they are not linearly independent, while for
DM that is an electroweak singlet, $d_\chi=1$, in addition the
operators $Q_5^{(6)}$ and $Q_6^{(6)}$ should be dropped from the
basis, since they both vanish in that case.

For a $d_\chi \geq 3$ DM multiplet there are seven operators that
involve Higgs currents,
\begin{align}
& Q_{11}^{(6)} = \left( \varphi^\dagger \varphi \right) \left( H^{\dagger}H\right)^2\,, 
& \qquad & Q_{12}^{(6)} = \left( \varphi^\dagger \tilde\tau^a \varphi \right) \left( H^{\dagger} \tau^a H\right)\left( H^{\dagger}H\right)\,,
 \label{eq:Q11:Q12:scal}\\
& Q_{13}^{(6)} = \Big( \varphi^\dagger i\stackrel{\leftrightarrow}{D}_\mu \varphi \Big) 
              \Big( H^{\dagger} i\stackrel{\leftrightarrow}{D^\mu} H\Big) \,,
& \qquad & Q_{14}^{(6)} = \Big( \varphi^\dagger i\stackrel{\leftrightarrow}{D^a}_\mu \varphi \Big) 
                      \Big( H^{\dagger} i\stackrel{\leftrightarrow}{D^{a,\mu}} H\Big)\,,
 \label{eq:Q13:Q14:scal}\\
& Q_{15}^{(6)} =  \Big( \varphi^\dagger \varphi \Big) 
              \Big( D^\mu H^{\dagger} D_\mu H\Big) \,,
& \qquad & Q_{16}^{(6)} = \Big( \varphi^\dagger \tilde\tau^a \varphi \Big) 
                       \Big( D^\mu H^{\dagger} \tau^a D_\mu H\Big)\,,
 \label{eq:Q15:Q16:scal}\\
&\Q_{17}^{(6)} = \left( \varphi^\dagger \tilde  \tau^{\{ab\}} \varphi \right) \left( H^{\dagger} \tau^a H\right)\left( H^{\dagger}\tau^bH\right) \,.
 \label{eq:Q17:scal}
\end{align}
The left-right covariant derivatives have been defined below
Eq.~\eqref{eq:dim6:Q16Q18}, with the appropriate representation of the
$\tau^a$ matrices understood. For DM that is an electroweak doublet,
$d_\chi=2$, the operator $Q_{17}^{(6)}$ should be dropped from the
basis. For singlet DM also the operators~$Q_{14}^{(6)}$
and~$Q_{16}^{(6)}$ should be dropped.

The Yukawa-like operators are, for both $d_\chi\geq 3$ and $d_\chi=2$,
\begin{align}
\label{eq:Q18i:Q19i}
&Q_{18,i}^{(6)} = \left( \varphi^\dagger \varphi \right) (\bar Q_L^i u_R^i \tilde H) \,,
&\qquad &Q_{19,i}^{(6)} = \left( \varphi^\dagger \tilde\tau^a \varphi \right)(\bar Q_L^i u_R^i \tau^a \tilde H)\,, 
\\
&Q_{20,i}^{(6)} = \left( \varphi^\dagger \varphi \right) (\bar Q_L^i d_R^i H) \,,
&\qquad &Q_{21,i}^{(6)} = \left( \varphi^\dagger \tilde\tau^a \varphi \right)(\bar Q_L^i d_R^i \tau^a H) \,,
\\
\label{eq:Q22i:Q23i}
&Q_{22,i}^{(6)} = \left( \varphi^\dagger \varphi \right) (\bar L_L^i \ell_R^i H) \,,
&\qquad &Q_{23,i}^{(6)} = \left( \varphi^\dagger \tilde\tau^a \varphi \right)(\bar L_L^i \ell_R^i \tau^a H)\,,
\end{align}
For singlet DM, $d_\chi=1$, the operators $Q_{19,i}^{(6)},
Q_{21,i}^{(6)}, Q_{23,i}^{(6)}$ do not appear in the basis.  Note that
the above operators, $Q_{18,i}^{(6)}, \ldots, Q_{23,i}^{(6)}$ are not
Hermitian, and their Wilson coefficients can be complex. Thus, in the
effective Lagrangian, Eq.~\eqref{eq:Lagr:EW:scal}, both the operators
$Q_{18,i}^{(6)}, \ldots, Q_{23,i}^{(6)}$ and their Hermitian
conjugates appear in the sum. Alternatively, one could use a set of 12
Hermitian operators (6 for $d_\chi=1$), i.e.,
$(Q_{a,i}^{(6)}+Q_{a,i}^{(6)\dagger})$, and
$i(Q_{a,i}^{(6)}-Q_{a,i}^{(6)\dagger})$, with real Wilson
coefficients.

Finally, the operators with fermion currents are, both for $d_\chi
\geq 3$ and $d_\chi=2$,
\begin{align}
&Q_{24,i}^{(6)} = \left( \varphi^\dagger i\!\!\stackrel{\leftrightarrow}{D}_\mu \varphi \right) 
               (\bar{Q}_L^i\gamma_\mu Q_L^i)\,, 
&\qquad &Q_{25,i}^{(6)} = \left( \varphi^\dagger i\!\!\stackrel{\leftrightarrow}{D}_\mu \tilde\tau^a \varphi \right) 
                        (\bar{Q}_L^i \gamma_\mu \tau^a Q_L^i)\,,
\\
&Q_{26,i}^{(6)} = \left( \varphi^\dagger i\!\!\stackrel{\leftrightarrow}{D}_\mu \varphi \right) 
                (\bar{u}_R^i\gamma_\mu u_R^i)\,,
&\qquad &Q_{27,i}^{(6)} = \left( \varphi^\dagger i\!\!\stackrel{\leftrightarrow}{D}_\mu \varphi \right) 
                        (\bar{d}_R^i\gamma_\mu d_R^i) \,,
\\
&Q_{28,i}^{(6)} =\left( \varphi^\dagger i\!\!\stackrel{\leftrightarrow}{D}_\mu \varphi \right) 
               (\bar{L}_L^i\gamma_\mu L_L^i)\,, 
&\qquad &Q_{29,i}^{(6)} = \left( \varphi^\dagger i\!\!\stackrel{\leftrightarrow}{D}_\mu \tilde\tau^a \varphi \right) 
                        (\bar{L}_L^i\gamma_\mu \tau^a L_L^i)\,,
\\
&Q_{30,i}^{(6)} = \left( \varphi^\dagger i\!\!\stackrel{\leftrightarrow}{D}_\mu \varphi \right) 
                (\bar{\ell}_R^i\gamma_\mu \ell_R^i)\,. 
 &\qquad 
\end{align}
For singlet DM, $d_\chi=1$, the operators $Q_{25,i}^{(6)}$ and
$Q_{29,i}^{(6)}$, should be dropped from the basis.

For completeness, we also display the dimension-seven operators for
scalar DM. For $d_\chi\geq 3$ there are three operators, all of which
violate lepton number by two units, $\Delta L=2$,
\begin{align}
Q^{(7)}_{1,i} &=\big(\varphi^\dagger \varphi\big)\big(\bar L_{L,i}^c \tilde H^* \tilde H^\dagger L_{L,i}\big)\,,
\\
Q^{(7)}_{2,i} &=\big(\varphi^\dagger \tilde \tau^a \varphi\big)\big(\bar L_{L,i}^c \tilde H^* \tilde H^\dagger \tau^a L_{L,i}\big)\,,
\\
Q^{(7)}_{3,i} &=\big(\varphi^\dagger \tilde \tau^{\{ab\}}\varphi\big)\big(\bar L_{L,i}^c (\tau^{a})^T \tilde H^* \tilde H^\dagger \tau^b L_{L,i}\big)\,.
\end{align}
For DM that is an electroweak doublet, $d_\chi=2$, the operator,
$Q^{(7)}_{3,i}$ is not linearly independent and should be dropped in
order to obtain a basis. For singlet DM both $Q^{(7)}_{2,i}$ and
$Q^{(7)}_{3,i}$ are zero.

This completes the construction of the EFT basis for operators with
mass dimension seven, for complex scalar DM. We have checked that the
basis is complete both with the algorithmic procedure outlined in
Appendix~\ref{sec:basis}, as well as with the conformal Hilbert series
method~\cite{Lehman:2015via, Henning:2015daa, Henning:2015alf}. 

For real scalar DM, the operators $Q^{(6)}_{13}$, $Q^{(6)}_{24,i}$,
$Q^{(6)}_{26,i}$, $Q^{(6)}_{27,i}$, $Q^{(6)}_{28,i}$, and
$Q^{(6)}_{30,i}$ vanish and should be dropped from the
basis. Note that the real scalar DM is in a real representation of $SU(2)$ and thus $d_\chi$ is necessarily odd.

\subsection{Matching below the electroweak scale, RG running, and nuclear response}

Below the electroweak scale the basis for DM interactions with the SM
fields contains 8 operators of dimension six (see
Ref.~\cite{Bishara:2017nnn}):
\begin{align}
&\mathcal{Q}_{1,f}^{(6)} = \left( \varphi^\dagger i \lrpartial^\mu \varphi \right) \left( \bar f \gamma_\mu f \right)\,, \label{eq:scal:Q1:Q2}
&\qquad  &\mathcal{Q}_{2,f}^{(6)} =  \left( \varphi^\dagger i \lrpartial^\mu \varphi \right) \left( \bar f \gamma_\mu \gamma_5 f \right)\,, \\
&\mathcal{Q}_{3,f}^{(6)} =  m_f \left( \varphi^\dagger \varphi \right) \left( \bar f f \right)\,, &\qquad  
&\mathcal{Q}_{4,f}^{(6)} =  m_f \left( \varphi^\dagger \varphi \right) \left( \bar f i \gamma_5 f \right)\,, \label{eq:scal:Q4:Q3}
\\
\label{eq:scal:Q5:Q6}
&\mathcal{Q}_5^{(6)} =  \frac{\alpha_s}{12\pi}\left( \varphi^\dagger \varphi \right) G_{\mu\nu}^aG^{\mu\nu}_a \,, 
&\qquad  &\mathcal{Q}_6^{(6)} =  \frac{\alpha_s}{8\pi}\left( \varphi^\dagger \varphi \right) \tilde{G}_{\mu\nu}^aG^{\mu\nu}_a\,, \\
&\mathcal{Q}_7^{(6)} =  \frac{\alpha}{12\pi} \left( \varphi^\dagger \varphi \right) F^{\mu\nu} F_{\mu\nu}\,, &\qquad
&  \mathcal{Q}_8^{(6)} =  \frac{\alpha}{8\pi} \left( \varphi^\dagger \varphi \right) F^{\mu\nu} \tilde F_{\mu\nu}\,.
\end{align}
Note that the operators $\mathcal{Q}_{1,f}^{(6)}$ and
$\mathcal{Q}_{2,f}^{(6)}$ vanish for real DM.

In the remainder of this section we present the explicit tree-level
matching conditions from the EFT above the electroweak scale to the
five-flavor EFT. The resulting Wilson coefficients can then be
evolved, using the RG equations, down to the hadronic scale $\mu \sim
2\,$GeV, where the matching onto the nuclear EFT is performed. The
details for the RG running and the subsequent matching to the nuclear
EFT for the above operators has been given in
Ref.~\cite{Bishara:2017nnn}, with the exception of the Rayleigh
operators $\Q_7^{(6)}, \Q_8^{(6)}$. Their nonperturbative matching is
directly analogous to the case of the fermionic operators
$Q_{11}^{(7)}\sim$ and $Q_{11}^{(7)}$, for which, however, only NDA
estimates are available at present (for the details see the previous
section).

Tree-level matching at the electroweak scale gives, for the Wilson
coefficients of the operators with the fermionic vector and axial
currents in Eq.~\eqref{eq:scal:Q1:Q2},
\begin{align}
\hat \C^{(6)}_{1(2),u_i} &= \frac{1}{2 \Lambda^2}
                        \left(\pm C^{(6)}_{24,i}  \mp \frac{Y_\chi}{4} C^{(6)}_{25,i} + C^{(6)}_{26,i}\right)\,, \\
\hat \C^{(6)}_{1(2),d_i} &= \frac{1}{2 \Lambda^2}
                        \left(  \pm C^{(6)}_{24,i}  \pm \frac{Y_\chi}{4} C^{(6)}_{25,i} + C^{(6)}_{27,i}\right)\,, \\
\hat \C^{(6)}_{1(2),\ell_i} &= \frac{1}{2 \Lambda^2}
                           \left(\pm C^{(6)}_{28,i}  \pm \frac{Y_\chi}{4} C^{(6)}_{29,i} + C^{(6)}_{30,i}\right)\,, 
\\
\hat \C^{(6)}_{1,\nu_i} &=-\hat \C^{(6)}_{2,\nu_i}
  = \frac{1}{2 \Lambda^2} \left(C^{(6)}_{28,i}  - \frac{Y_\chi}{4} C^{(6)}_{29,i} \right)\,. 
\end{align}
The operators involving scalar currents, Eq.~\eqref{eq:scal:Q4:Q3},
obtain the following contributions:
\begin{align}
\begin{split}
\hat \C^{(6)}_{3,u_i}
 &= \frac{1}{\Lambda^2}\left[\frac{v_{\rm EW}}{\sqrt{2} m_{u_i}}
    \text{Re} \Big( C^{(6)}_{18,i} - \frac{Y_\chi}{4} C^{(6)}_{19,i}\Big)
   -\frac{v_{\rm EW}^2}{M_h^2}
    \Big(C^{(6)}_{11} + \frac{Y_\chi}{4} C^{(6)}_{12} + \frac{Y_\chi^2}{8} C^{(6)}_{17} \Big)\right]\\
 & \quad + \frac{1}{4M_h^2} \Big( \lambda_{\varphi H}+\frac{Y_\chi}{4}\lambda_{\varphi H}' \Big) + \cdots\,,
\end{split}
\\
\hat \C^{(6)}_{4,u_i}
 & = \frac{1}{\Lambda^2} \frac{v_{\rm EW}}{\sqrt{2} m_{u_i}}
     \text{Im} \left( C^{(6)}_{18,i} - \frac{Y_\chi}{4} C^{(6)}_{19,i} \right)\,
\end{align}
for up-type quarks,
\begin{align}
\begin{split}
\hat \C^{(6)}_{3,d_i}
 &= \frac{1}{\Lambda^3}\left[\frac{v_{\rm EW}}{\sqrt{2} m_{d_i}}
    \text{Re} \Big( C^{(6)}_{20,i} + \frac{Y_\chi}{4} C^{(6)}_{21,i}\Big)
   -\frac{v_{\rm EW}^2}{M_h^2}
    \Big(C^{(6)}_{11}+\frac{Y_\chi}{4} C^{(6)}_{12} + \frac{Y_\chi^2}{8} C^{(6)}_{17} \Big)\right]\\
 & \quad + \frac{1}{4M_h^2} \Big( \lambda_{\varphi H}+\frac{Y_\chi}{4}\lambda_{\varphi H}' \Big) +\cdots\,,
\end{split}
\\
\hat \C^{(6)}_{4,d_i}
 & = \frac{1}{\Lambda^3}\frac{v_{\rm EW}}{\sqrt{2} m_{d_i}}
     \text{Im} \left( C^{(6)}_{20,i} + \frac{Y_\chi}{4} C^{(6)}_{21,i} \right)\,
\end{align}
for down-type quarks, and
\begin{align}
\begin{split}
\hat \C^{(6)}_{3,\ell_i}
 &= \frac{1}{\Lambda^3}\left[\frac{v_{\rm EW}}{\sqrt{2} m_{\ell_i}}
    \text{Re} \Big( C^{(6)}_{22,i} + \frac{Y_\chi}{4} C^{(6)}_{23,i}\Big)
   -\frac{v_{\rm EW}^2}{M_h^2}
    \Big(C^{(6)}_{11}+\frac{Y_\chi}{4} C^{(6)}_{12} + \frac{Y_\chi^2}{8} C^{(6)}_{17} \Big)\right]\\
 & \quad + \frac{1}{4M_h^2} \Big( \lambda_{\varphi H}+\frac{Y_\chi}{4}\lambda_{\varphi H}' \Big) +\cdots\,,
\end{split}
\\
\hat \C^{(6)}_{4,\ell_i}
 & = \frac{1}{\Lambda^3}\frac{v_{\rm EW}}{\sqrt{2} m_{\ell_i}}
     \text{Im} \left( C^{(6)}_{22,i} + \frac{Y_\chi}{4} C^{(6)}_{23,i} \right)\,
\end{align}
for charged leptons. The matching for the gluonic operators,
Eq.~\eqref{eq:scal:Q5:Q6}, gives
\begin{equation}
\hat \C^{(6)}_{5(6)} = \frac{1}{\Lambda^2} C^{(6)}_{1(2)}\,,
\end{equation} 
while for the Rayleigh operators we have
\begin{equation}
\hat \C^{(6)}_{7(8)} = \frac{1}{\Lambda^2}
\left(C^{(6)}_{3(4)}+\frac{Y_\chi}{2}C^{(6)}_{5(6)} +C^{(6)}_{7(8)} +
\frac{Y_\chi^2}{2} C^{(6)}_{9(10)}\right)\,.
\end{equation} 
Note that, at tree level, the two-Higgs operators with derivatives in
Eqs.~\eqref{eq:Q13:Q14:scal} do not give contributions to the
dimension-six operators below the electroweak scale.

\section{Conclusions}
\label{sec:conclusions}
We presented a complete operator basis for an Effective Field Theory
(EFT) that describes the interactions between DM bilinears and the SM
fields, up to and including mass dimension seven, both for fermionic
and scalar DM. For Dirac fermion and complex scalar DM we restricted
the discussion to the case where the EFT obeys a dark global $U(1)_D$
symmetry, under which the DM field carries a nonzero charge. For DM
that is a Dirac fermion in a general electroweak $SU(2)$ multiplet of
dimensionality $d_\chi\geq 3$, the basis for an EFT valid above the
electroweak symmetry breaking scale contains 100 Hermitian
dimension-seven operators. We have shown that such a basis is
complete. A reduction of the basis occurs, if DM is an electroweak
doublet or a singlet, or if DM is a Majorana fermion, as discussed in
detail in the main text.

We provided both the tree-level matching at the electroweak scale onto
an EFT with five quark flavors, valid for $m_b\lesssim \mu \lesssim
m_W$, as well as the renormalization group running down to an EFT with
four quark flavors, valid for $\mu \lesssim m_b$, and down to three
quark flavor EFT, where also the charm quark is integrated out. We
also presented the nonperturbative matching onto the nuclear EFT at
leading order in chiral counting.

There are a number of UV models of DM that give the first nonzero
contributions only at dimension seven. For instance, if DM is a
Majorana fermion, a singlet under the electroweak gauge group, but
couples to electroweakly charged heavy states, the leading interaction
at low energies can well be the Rayleigh operators,
Eqs.~\eqref{eq:Q7:11} and~\eqref{eq:Q7:13} (see
Ref.~\cite{Weiner:2012gm} for explicit models). Another popular set of
UV models that give dimension-seven operators as the leading effect
has DM coupling to the visible sector through scalar mediators. The
scalar mediator exchanges in the $s$-channel (for DM annihilation)
lead to the scalar operators in Eq.~\eqref{eq:Q7:5:6}, while
$t$-channel scalar mediator exchanges lead to a linear combination of
scalar and tensor operators, Eqs.~\eqref{eq:Q7:5:6}
and~\eqref{eq:Q7:9:10}, after Fierz transformations.

The EFT we presented is valid as long as the mediators are heavier
than the momenta exchanges in the physical process one is
considering. For direct detection the typical momentum exchange is
below 200 MeV. This means that the direct detection processes for a
large class of the UV theories of DM would be described by the EFT
discussed in this paper. Still, one can think of ways to extend the
framework further. For instance, one could extend the basis to include
operators with several DM multiplets, or to relax the assumption of
global dark $U(1)_D$. We leave these for future work.

\acknowledgments We thank Fady Bishara for useful discussions, and
Zach Polonsky and Jos\'e Santiago for pointing out two missing
operators. JZ and MT acknowledge support in part by the U.S. National
Science Foundation under CAREER Grant PHY-1151392 and by the DOE grant
de-sc0011784.

\appendix

\section{Construction of the operator basis}\label{sec:basis} 

Here we briefly describe the algorithmic procedure that was used to
construct the operator basis (see also Ref.~\cite{Grzadkowski:2010es,
  Gripaios:2016xuo}). We concentrate on the EFT valid above the
electroweak breaking scale, so that the DM furnishes a general
representation of the $SU(2)$ gauge group. The construction of the
operator basis below the electroweak scale proceeds in an analogous,
albeit simpler way.

We start by writing all possible gauge- and Lorentz-invariant
contractions of two DM fields with any number of SM fields to
construct all possible operators of a given mass dimension. We then
use various relations to eliminate redundant operators. Note that we
do not include in our basis the operators that involve
charge-conjugated dark-matter fields. For Majorana DM these do not
lead to any additional operators. However, for Dirac DM one could
obtain additional operators, depending on the choice of the DM
hypercharge, $Y_\chi$, which does not cancel in the $\bar \chi^c
\Gamma \chi$ current (unlike the $\bar \chi \Gamma \chi$ currents).
For example, for $Y_\chi=-1/2$ the dimension-seven operator
\begin{equation}
(\bar{\chi}^c \chi) (\bar{L}E^c\tilde {H})
\end{equation}
is gauge invariant and could be added to the operator basis. Above,
the charge conjugation of the DM field $\chi$ includes an appropriate
transformation of its $SU(2)$ representation.

In the remainder of this section we will discuss all relations that
were exploited to remove linearly dependent operators. For more
details on the actual implementation into {\tt C++} code, see
Ref.~\cite{AWGD:master-thesis}.

\paragraph*{Permutations of identical fields and index symmetries.}
We explicitly check the disparity of operators under simultaneous
permutations of (anti-)symmetric indices and index sets of identical
fields, keeping track of fermionic signs. This also allows us to
eliminate operators that vanish already via antisymmetry of indices.

\paragraph*{Chirality of standard model fermions.}
We eliminate all operators that vanish via the chirality of the
respective fields.

\paragraph*{De-facto symmetry of covariant derivatives.}
Pairs of covariant derivatives acting on the same field can be
decomposed into parts that are symmetric and antisymmetric,
respectively. Since the antisymmetric part is proportional to the
commutator $[D_\mu, D_\nu]$, which is in turn given in terms of
gauge field strength tensors, it is already covered by operators with
a lower count of covariant derivatives. 

\paragraph*{Integration by parts.}
Since total derivatives leave the action invariant, expanding them
using the Leibniz rule yields additional relations between our
operators. 

\paragraph*{Fierz identities.}
The general form of the well-known Fierz identities (see, for
instance, Ref.~\cite{Nishi:2004st, Borodulin:2017pwh}) allows us to
disregard contractions that are not in a standard order (though this
is arbitrarily chosen). Any operator obtained by reordering the
pattern of contractions will be a linear combination of the previous
operators. This argument also applies to more than two bilinears,
since a general permutation of fields can be expressed as a number of
transpositions, for which the usual formula holds.

While the Fierz relations can be used to move all bilinears into a
standard ordering, there are two cases where they imply additional
relations. First, if two fields in the bilinears are identical,
application of a Fierz identity directly relates terms that are
already in the standard order. Second, if an expression in a
non-standard order vanishes, for instance, by chirality, the equation
fo motion (EOM), or integration-by-part identities, then the
equivalent expression in the standard ordering must vanish too,
possibly yielding a new, independent relation.

\paragraph*{Schouten identities.}
These follow from the fact that there is no totally antisymmetric
tensor with more indices than the vector space dimension, other than
the zero tensor. In our case, antisymmetrizing an object with more
than four Lorentz indices yields zero. For instance, antisymmetrizing
the indices of the Levi-Civita symbol together with $\mu$ in
\begin{equation}
(\bar{\chi}\sigma_{\mu\nu}\chi) B^\mu_{\phantom{\mu}\sigma}B_{\alpha\beta}\varepsilon^{\nu\sigma\alpha\beta}\,,
\end{equation}
and dropping terms that vanish by symmetry, implies that the whole
operator vanishes. Moreover, Schouten identities can also yield
relations like
\begin{equation}
\bar{\chi}\sigma_{\mu\nu}\tau^a\chi\; W^{b\phantom{\sigma}\nu}_{\phantom{b}\sigma} W^c_{\alpha\beta}\;\varepsilon^{abc}\varepsilon^{\mu\sigma\alpha\beta}
=
\bar{\chi}\sigma_{\mu\nu}\tau^a\chi\; W^{b\phantom{\sigma}\alpha}_{\phantom{b}\sigma} W^c_{\alpha\beta}\;\varepsilon^{abc}\varepsilon^{\mu\sigma\nu\beta}\,,
\end{equation}
which follows from antisymmetrizing the Levi-Civita symbol, together
with the index $\nu$, on the left-hand side.

\paragraph*{Bianchi identity.}
The Bianchi identity for any gauge field strength tensor
$G^a_{\mu\nu}$ is given by
\begin{equation}
D_\mu G^a_{\nu\rho}+ D_\nu G^a_{\rho\mu}+ D_\rho G^a_{\mu\nu} =0\,.
\end{equation}
In particular, this implies that the covariant derivative of the dual
tensor vanishes,
\begin{equation}
D^\mu \tilde{G}^a_{\mu\nu} = 0\,.
\end{equation}

\paragraph*{Additional reductions for special representations of $SU(2)$.}\label{par:additionalReductions}
Our operator basis is valid for a general irreducible representation
of $SU(2)$ that the dark matter field $\chi$ furnishes. Additional
relations, however, can make some operators redundant for specific
representations. For the fundamental representation, $d_\chi=2$, some
of the operators in the general basis are linearly dependent due to
the completeness relation, Eq.~\eqref{eq:completeness}. The reduced
set of operators is given explicitly in the main text.  For the
trivial representation of $SU(2)$, $d_\chi=1$, one simply drops all
the operators with an $SU(2)$ generator in the DM current.

\paragraph*{Additional reduction for the Majorana / real scalar case.}\label{par:additionalReductionsMajorana}
For every Majorana DM bilinear we can use the Majorana condition to
replace $\chi$ with its charge conjugate (see, e.g.,
\cite{Dreiner:2008tw}).  A dark matter bilinear that is odd under
charge symmetry can then be dropped from the operator basis in the
case of Majorana DM, eliminating DM vector and tensor currents (the
covariant derivatives acting within the bilinear can change this). The
list of these operators is given explicitly at the end of
Section~\ref{sec:EFT:EWK}. In complete analogy, the operators listed
at the end of Section~\ref{sec:scal:basis:ew} vanish for real scalar
DM.

\paragraph*{Equations of motion.}
The EOM of the SM and DM fields imply relations between operators that
can be used to reduce the operator basis~\cite{Simma:1993ky}. For
scalar fields and field strength tensors, the reduction using the EOM
is fairly straightforward. For Dirac fields, however, the Clifford
algebra complicates matters. To deal with this, we first construct a
new set of EOM-vanishing operators by finding all operators of the
appropriate mass dimension with any of the following matrix
expressions in Dirac field bilinears,
\begin{equation}
\{ \gamma^\mu\,, \quad \gamma^\mu\gamma_5\,, \quad
\gamma^\mu\gamma^\nu\,, \quad \gamma^\mu\gamma^\nu\gamma_5\,, \quad
\gamma^\mu\gamma^\nu\gamma^\rho\,, \quad
\gamma^\mu\gamma^\nu\gamma^\rho\gamma_5 \}\,,
\end{equation}
where either the rightmost or leftmost Lorentz index is contracted
with a covariant derivative acting on the right or left Dirac field,
respectively. We then translate each of these terms into relations
between elements of our full operator list by exploiting the
identities
\begin{align} 
\gamma_\mu \gamma_\nu &= \eta_{\mu\nu}-i\sigma_{\mu\nu}\,, \label{eq:eom-gamma2} \\ 
\gamma_\mu \gamma_\nu\gamma_5 &= \eta_{\mu\nu}\gamma_5-i\sigma_{\mu\nu}\gamma_5\,, \label{eq:eom-gamma25} \\ 
\gamma_\mu \gamma_\nu \gamma_\rho &= \eta_{\mu\nu} \gamma_\rho + \eta_{\nu\rho}\gamma_\mu - \eta_{\mu\rho} \gamma_\nu - i \varepsilon_{\sigma\mu\nu\rho} \gamma^\sigma \gamma_5\,, \label{eq:eom-gamma3} \\
\gamma_\mu \gamma_\nu \gamma_\rho\gamma_5 &= \eta_{\mu\nu} \gamma_\rho\gamma_5 + \eta_{\nu\rho}\gamma_\mu\gamma_5 - \eta_{\mu\rho} \gamma_\nu\gamma_5 - i \varepsilon_{\sigma\mu\nu\rho}\gamma^\sigma\,, \label{eq:eom-gamma35}
\end{align}
where the chirality projector of the SM fields is used to absorb
$\gamma_5$, if possible. If there already exists another Lorentz
Levi-Civita symbol in the remainder of the operator, their product
gets simplified as described above.

This procedure, together with the previously discussed treatment of
partial integration and index symmetries, captures all relations
implied by equations of motion. To see this, consider a completely
general EOM-vanishing operator
\begin{equation}
  \big( \bar{\Psi} \Gamma \gamma^\mu D_\mu \Psi \big) X \,,
\end{equation}
where $\Psi$ is any Dirac field, $\Gamma$ is any sum of products of
Dirac matrices, possibly contracted with other expressions in the
remainder of the operator, which is denoted by $X$. This operator,
when expressed in our basis, does not induce any relations beyond the
ones constructed above. Eqs.~\eqref{eq:eom-gamma3}
and~\eqref{eq:eom-gamma35} can be repeatedly applied within each term
of $\Gamma$ to reduce it to a sum of terms with a maximum of two gamma
matrices (and possibly one $\gamma_5$, which can always be
anticommuted to the rightmost position and eliminated if an even
number of them exists). We have thus expressed the above operator as a
linear combination of the vanishing operators already constructed. The
case where the covariant derivative acts on the left, or if additional
derivatives appear, can be treated analogously.

Finally, as a consistency check of the constructed basis, we compared
it with the operator counts for a given field content that were
derived using the conformal Hilbert series
method~\cite{Lehman:2015via, Henning:2015daa, Henning:2015alf}. This
approach allows for a systematic group-theoretic treatment of
equations of motion and inclusion of integration-by-parts
identities. It calculates an object called the Hilbert series, given
by
\begin{equation}
H(\mathcal{D}, \phi_1, \dots, \phi_N) = \sum\limits_{k,r_1,\dots,r_N} \, c_{k,r_1,\dots,r_N}\; \phi^{r_1}_1 \dots \phi^{r_N}_N \;\mathcal{D}^k\,,
\end{equation}
where $\phi_i$ and $\mathcal{D}$ are complex numbers that stand in for
the fields of the theory and the covariant derivative. The
coefficients $c_{k,r_1,\dots,r_N}$ are the desired operator counts.

The Mathematica package provided in Ref.~\cite{Henning:2015alf}, with
the addition of the fermionic or scalar DM field in an $SU(2)$
multiplet, gives the total number of independent, Lorentz- and
gauge-invariant operators of a certain mass dimension, as well as
their field content. We have verified that the number of operators
within each operator class obtained in this way exactly matches the
number of operators constructed using our algorithmic procedure.

\section{Non-relativistic reduction of fermion bilinears}
\label{sec:NR:app}
To match onto an EFT with nonrelativistic nucleons, we first introduce
a Heavy DM Effective Theory (HDMET) field (for further details see
Ref.~\cite{Bishara:2017pfq}),
\begin{equation}\label{eq:chi-field-def}
  \chi (x) = e^{-i m_\chi v \cdot x} \big( \chi_v (x) + X_v (x) \big) \,,
\end{equation}
where 
\begin{equation}
  \chi_v (x) =  e^{i m_\chi v \cdot x} \frac{1 + {\slashed v}}{2} \chi
  (x) \,, \qquad X_v (x) = e^{i m_\chi v \cdot x} \frac{1 - \slashed
    v}{2} \chi (x) \,.
\end{equation}
Integrating out the antiparticle modes, we obtain the tree-level
relation~\cite{Neubert:1993mb}
\begin{equation}
\chi=e^{-i m_\chi v \cdot x} \Big(1 +\frac{i \slashed
  \partial_\perp}{i v\cdot \partial+2 m_\chi-i \epsilon}\Big)
\chi_v\,,\label{eq:chi:rel}
\end{equation}
where $\gamma_\perp^\mu = \gamma^\mu - v^\mu \slashed{v}$, 
so that we obtain the conventional ``NRQED'' type Lagrangian,
\begin{equation}\label{eq:L:NRQED}
{\cal L}_{\rm HDMET}=\chi_v^\dagger \Big( i v\cdot \partial+\frac{ (i \partial_\perp)^2}{2 m_\chi}+ \cdots \Big)\chi_v\,,
\end{equation}
where $\partial_\perp^\mu=\partial^\mu-v\cdot \partial\, v^\mu$.
 Using the
replacement~\eqref{eq:chi:rel} and applying the equation of motion derived from
Eq.~\eqref{eq:L:NRQED}, we obtain the following nonrelativistic DM
currents:
\begin{align}
\label{eq:HDMETlimit:dscalar}
& \big( \bar \chi i \lrpartial_\mu \chi \big)
  \to 2 m_\chi v_\mu \bar
\chi_v \chi_v + \bar \chi_v i \lrpartial_\mu \chi_v + \ldots\,, \\
\label{eq:HDMETlimit:dpscalar}
& 
\big( \bar \chi i \gamma_5 i\lrpartial_\mu
\chi \big) \to 2 v_\mu \partial_\rho \big(\bar \chi_v S_\chi^\rho \chi_v
\big) + \frac{1}{ m_\chi} \partial_\rho \big(\bar \chi_v S_\chi^\rho i
\lrpartial_\mu \chi_v \big) + \ldots\,,
\\
\begin{split}\label{eq:dtensorDM:expand}
& 
\partial_\mu \big( \bar \chi \sigma^{\mu\nu} \chi \big) \to -2
\epsilon^{\alpha\beta\nu\rho} v_\alpha \partial_\beta \big( \bar
\chi_v S_{\chi,\rho} \chi_v \big) \\ & \qquad \qquad \qquad + \frac{v^\nu}{2m_\chi}
\Big( \partial^2 \big( \bar\chi_v \chi_v \big) - 2
\epsilon^{\alpha\beta\rho\sigma} v_\alpha \partial_\rho \big( \bar
\chi_v S_{\chi,\sigma} i \lrpartial_\beta \chi_v \big) \Big) +
\ldots\,,
\end{split}
\\
\begin{split}\label{eq:daxialtensorDM:expand}
& \partial_\mu \big( \bar \chi \sigma^{\mu\nu} i\gamma_5 \chi \big)
\to 2 v^\nu \partial_\rho \big( \bar \chi_v S_\chi^\rho \chi_v \big) 
+ \frac{i}{m_\chi} \partial_\rho \big( \bar \chi_v S_\chi^{\rho}
\lrpartial^{\nu} \chi_v \big)+ \cdots\,,
\end{split}
\end{align}
that enter the nonrelativistic reduction of the new dimension-seven
operators, Eqs.~\eqref{eq:4Fd7:3}-\eqref{eq:4Fd7:4}. The ellipses
denote higher orders in the expansion in $1/m_\chi$. The remaining
nonrelativistic currents, as well as the corresponding products with
the expanded nuclear form factors, have already been presented in
Ref.~\cite{Bishara:2017pfq}.

\bibliography{DMDIM7}

\begin{thebibliography}{46}
\expandafter\ifx\csname natexlab\endcsname\relax\def\natexlab#1{#1}\fi
\expandafter\ifx\csname bibnamefont\endcsname\relax
  \def\bibnamefont#1{#1}\fi
\expandafter\ifx\csname bibfnamefont\endcsname\relax
  \def\bibfnamefont#1{#1}\fi
\expandafter\ifx\csname citenamefont\endcsname\relax
  \def\citenamefont#1{#1}\fi
\expandafter\ifx\csname url\endcsname\relax
  \def\url#1{\texttt{#1}}\fi
\expandafter\ifx\csname urlprefix\endcsname\relax\def\urlprefix{URL }\fi
\providecommand{\bibinfo}[2]{#2}
\providecommand{\eprint}[2][]{\url{#2}}

\bibitem[{\citenamefont{Bishara
  et~al.}(2017{\natexlab{a}})\citenamefont{Bishara, Brod, Grinstein, and
  Zupan}}]{Bishara:2016hek}
\bibinfo{author}{\bibfnamefont{F.}~\bibnamefont{Bishara}},
  \bibinfo{author}{\bibfnamefont{J.}~\bibnamefont{Brod}},
  \bibinfo{author}{\bibfnamefont{B.}~\bibnamefont{Grinstein}},
  \bibnamefont{and} \bibinfo{author}{\bibfnamefont{J.}~\bibnamefont{Zupan}},
  \bibinfo{journal}{JCAP} \textbf{\bibinfo{volume}{1702}}, \bibinfo{pages}{009}
  (\bibinfo{year}{2017}{\natexlab{a}}), \eprint{1611.00368}.

\bibitem[{\citenamefont{Fan et~al.}(2010)\citenamefont{Fan, Reece, and
  Wang}}]{Fan:2010gt}
\bibinfo{author}{\bibfnamefont{J.}~\bibnamefont{Fan}},
  \bibinfo{author}{\bibfnamefont{M.}~\bibnamefont{Reece}}, \bibnamefont{and}
  \bibinfo{author}{\bibfnamefont{L.-T.} \bibnamefont{Wang}},
  \bibinfo{journal}{JCAP} \textbf{\bibinfo{volume}{1011}}, \bibinfo{pages}{042}
  (\bibinfo{year}{2010}), \eprint{1008.1591}.

\bibitem[{\citenamefont{Fitzpatrick et~al.}(2013)\citenamefont{Fitzpatrick,
  Haxton, Katz, Lubbers, and Xu}}]{Fitzpatrick:2012ix}
\bibinfo{author}{\bibfnamefont{A.~L.} \bibnamefont{Fitzpatrick}},
  \bibinfo{author}{\bibfnamefont{W.}~\bibnamefont{Haxton}},
  \bibinfo{author}{\bibfnamefont{E.}~\bibnamefont{Katz}},
  \bibinfo{author}{\bibfnamefont{N.}~\bibnamefont{Lubbers}}, \bibnamefont{and}
  \bibinfo{author}{\bibfnamefont{Y.}~\bibnamefont{Xu}}, \bibinfo{journal}{JCAP}
  \textbf{\bibinfo{volume}{1302}}, \bibinfo{pages}{004} (\bibinfo{year}{2013}),
  \eprint{1203.3542}.

\bibitem[{\citenamefont{Fitzpatrick et~al.}(2012)\citenamefont{Fitzpatrick,
  Haxton, Katz, Lubbers, and Xu}}]{Fitzpatrick:2012ib}
\bibinfo{author}{\bibfnamefont{A.~L.} \bibnamefont{Fitzpatrick}},
  \bibinfo{author}{\bibfnamefont{W.}~\bibnamefont{Haxton}},
  \bibinfo{author}{\bibfnamefont{E.}~\bibnamefont{Katz}},
  \bibinfo{author}{\bibfnamefont{N.}~\bibnamefont{Lubbers}}, \bibnamefont{and}
  \bibinfo{author}{\bibfnamefont{Y.}~\bibnamefont{Xu}} (\bibinfo{year}{2012}),
  \eprint{1211.2818}.

\bibitem[{\citenamefont{Anand et~al.}(2014)\citenamefont{Anand, Fitzpatrick,
  and Haxton}}]{Anand:2013yka}
\bibinfo{author}{\bibfnamefont{N.}~\bibnamefont{Anand}},
  \bibinfo{author}{\bibfnamefont{A.~L.} \bibnamefont{Fitzpatrick}},
  \bibnamefont{and} \bibinfo{author}{\bibfnamefont{W.~C.}
  \bibnamefont{Haxton}}, \bibinfo{journal}{Phys. Rev.}
  \textbf{\bibinfo{volume}{C89}}, \bibinfo{pages}{065501}
  (\bibinfo{year}{2014}), \eprint{1308.6288}.

\bibitem[{\citenamefont{Cirelli et~al.}(2013)\citenamefont{Cirelli, Del~Nobile,
  and Panci}}]{DelNobile:2013sia}
\bibinfo{author}{\bibfnamefont{M.}~\bibnamefont{Cirelli}},
  \bibinfo{author}{\bibfnamefont{E.}~\bibnamefont{Del~Nobile}},
  \bibnamefont{and} \bibinfo{author}{\bibfnamefont{P.}~\bibnamefont{Panci}},
  \bibinfo{journal}{JCAP} \textbf{\bibinfo{volume}{1310}}, \bibinfo{pages}{019}
  (\bibinfo{year}{2013}), \eprint{1307.5955}.

\bibitem[{\citenamefont{Barello et~al.}(2014)\citenamefont{Barello, Chang, and
  Newby}}]{Barello:2014uda}
\bibinfo{author}{\bibfnamefont{G.}~\bibnamefont{Barello}},
  \bibinfo{author}{\bibfnamefont{S.}~\bibnamefont{Chang}}, \bibnamefont{and}
  \bibinfo{author}{\bibfnamefont{C.~A.} \bibnamefont{Newby}},
  \bibinfo{journal}{Phys. Rev.} \textbf{\bibinfo{volume}{D90}},
  \bibinfo{pages}{094027} (\bibinfo{year}{2014}), \eprint{1409.0536}.

\bibitem[{\citenamefont{Hill and Solon}(2015)}]{Hill:2014yxa}
\bibinfo{author}{\bibfnamefont{R.~J.} \bibnamefont{Hill}} \bibnamefont{and}
  \bibinfo{author}{\bibfnamefont{M.~P.} \bibnamefont{Solon}},
  \bibinfo{journal}{Phys.Rev.} \textbf{\bibinfo{volume}{D91}},
  \bibinfo{pages}{043505} (\bibinfo{year}{2015}), \eprint{1409.8290}.

\bibitem[{\citenamefont{Catena and Gondolo}(2014)}]{Catena:2014uqa}
\bibinfo{author}{\bibfnamefont{R.}~\bibnamefont{Catena}} \bibnamefont{and}
  \bibinfo{author}{\bibfnamefont{P.}~\bibnamefont{Gondolo}},
  \bibinfo{journal}{JCAP} \textbf{\bibinfo{volume}{1409}}, \bibinfo{pages}{045}
  (\bibinfo{year}{2014}), \eprint{1405.2637}.

\bibitem[{\citenamefont{Kopp et~al.}(2010)\citenamefont{Kopp, Schwetz, and
  Zupan}}]{Kopp:2009qt}
\bibinfo{author}{\bibfnamefont{J.}~\bibnamefont{Kopp}},
  \bibinfo{author}{\bibfnamefont{T.}~\bibnamefont{Schwetz}}, \bibnamefont{and}
  \bibinfo{author}{\bibfnamefont{J.}~\bibnamefont{Zupan}},
  \bibinfo{journal}{JCAP} \textbf{\bibinfo{volume}{1002}}, \bibinfo{pages}{014}
  (\bibinfo{year}{2010}), \eprint{0912.4264}.

\bibitem[{\citenamefont{Hill and Solon}(2014)}]{Hill:2013hoa}
\bibinfo{author}{\bibfnamefont{R.~J.} \bibnamefont{Hill}} \bibnamefont{and}
  \bibinfo{author}{\bibfnamefont{M.~P.} \bibnamefont{Solon}},
  \bibinfo{journal}{Phys. Rev. Lett.} \textbf{\bibinfo{volume}{112}},
  \bibinfo{pages}{211602} (\bibinfo{year}{2014}), \eprint{1309.4092}.

\bibitem[{\citenamefont{Hill and Solon}(2012)}]{Hill:2011be}
\bibinfo{author}{\bibfnamefont{R.~J.} \bibnamefont{Hill}} \bibnamefont{and}
  \bibinfo{author}{\bibfnamefont{M.~P.} \bibnamefont{Solon}},
  \bibinfo{journal}{Phys.Lett.} \textbf{\bibinfo{volume}{B707}},
  \bibinfo{pages}{539} (\bibinfo{year}{2012}), \eprint{1111.0016}.

\bibitem[{\citenamefont{Kurylov and Kamionkowski}(2004)}]{Kurylov:2003ra}
\bibinfo{author}{\bibfnamefont{A.}~\bibnamefont{Kurylov}} \bibnamefont{and}
  \bibinfo{author}{\bibfnamefont{M.}~\bibnamefont{Kamionkowski}},
  \bibinfo{journal}{Phys. Rev.} \textbf{\bibinfo{volume}{D69}},
  \bibinfo{pages}{063503} (\bibinfo{year}{2004}), \eprint{hep-ph/0307185}.

\bibitem[{\citenamefont{Pospelov and ter Veldhuis}(2000)}]{Pospelov:2000bq}
\bibinfo{author}{\bibfnamefont{M.}~\bibnamefont{Pospelov}} \bibnamefont{and}
  \bibinfo{author}{\bibfnamefont{T.}~\bibnamefont{ter Veldhuis}},
  \bibinfo{journal}{Phys. Lett.} \textbf{\bibinfo{volume}{B480}},
  \bibinfo{pages}{181} (\bibinfo{year}{2000}), \eprint{hep-ph/0003010}.

\bibitem[{\citenamefont{Bagnasco et~al.}(1994)\citenamefont{Bagnasco, Dine, and
  Thomas}}]{Bagnasco:1993st}
\bibinfo{author}{\bibfnamefont{J.}~\bibnamefont{Bagnasco}},
  \bibinfo{author}{\bibfnamefont{M.}~\bibnamefont{Dine}}, \bibnamefont{and}
  \bibinfo{author}{\bibfnamefont{S.~D.} \bibnamefont{Thomas}},
  \bibinfo{journal}{Phys. Lett.} \textbf{\bibinfo{volume}{B320}},
  \bibinfo{pages}{99} (\bibinfo{year}{1994}), \eprint{hep-ph/9310290}.

\bibitem[{\citenamefont{Cirigliano et~al.}(2012)\citenamefont{Cirigliano,
  Graesser, and Ovanesyan}}]{Cirigliano:2012pq}
\bibinfo{author}{\bibfnamefont{V.}~\bibnamefont{Cirigliano}},
  \bibinfo{author}{\bibfnamefont{M.~L.} \bibnamefont{Graesser}},
  \bibnamefont{and}
  \bibinfo{author}{\bibfnamefont{G.}~\bibnamefont{Ovanesyan}},
  \bibinfo{journal}{JHEP} \textbf{\bibinfo{volume}{10}}, \bibinfo{pages}{025}
  (\bibinfo{year}{2012}), \eprint{1205.2695}.

\bibitem[{\citenamefont{Hoferichter et~al.}(2015)\citenamefont{Hoferichter,
  Klos, and Schwenk}}]{Hoferichter:2015ipa}
\bibinfo{author}{\bibfnamefont{M.}~\bibnamefont{Hoferichter}},
  \bibinfo{author}{\bibfnamefont{P.}~\bibnamefont{Klos}}, \bibnamefont{and}
  \bibinfo{author}{\bibfnamefont{A.}~\bibnamefont{Schwenk}},
  \bibinfo{journal}{Phys. Lett.} \textbf{\bibinfo{volume}{B746}},
  \bibinfo{pages}{410} (\bibinfo{year}{2015}), \eprint{1503.04811}.

\bibitem[{\citenamefont{Hoferichter et~al.}(2016)\citenamefont{Hoferichter,
  Klos, Men\'{e}ndez, and Schwenk}}]{Hoferichter:2016nvd}
\bibinfo{author}{\bibfnamefont{M.}~\bibnamefont{Hoferichter}},
  \bibinfo{author}{\bibfnamefont{P.}~\bibnamefont{Klos}},
  \bibinfo{author}{\bibfnamefont{J.}~\bibnamefont{Men\'{e}ndez}},
  \bibnamefont{and} \bibinfo{author}{\bibfnamefont{A.}~\bibnamefont{Schwenk}},
  \bibinfo{journal}{Phys. Rev.} \textbf{\bibinfo{volume}{D94}},
  \bibinfo{pages}{063505} (\bibinfo{year}{2016}), \eprint{1605.08043}.

\bibitem[{\citenamefont{Bishara
  et~al.}(2017{\natexlab{b}})\citenamefont{Bishara, Brod, Grinstein, and
  Zupan}}]{Bishara:2017pfq}
\bibinfo{author}{\bibfnamefont{F.}~\bibnamefont{Bishara}},
  \bibinfo{author}{\bibfnamefont{J.}~\bibnamefont{Brod}},
  \bibinfo{author}{\bibfnamefont{B.}~\bibnamefont{Grinstein}},
  \bibnamefont{and} \bibinfo{author}{\bibfnamefont{J.}~\bibnamefont{Zupan}},
  \bibinfo{journal}{JHEP} \textbf{\bibinfo{volume}{11}}, \bibinfo{pages}{059}
  (\bibinfo{year}{2017}{\natexlab{b}}), \eprint{1707.06998}.

\bibitem[{\citenamefont{Menendez et~al.}(2012)\citenamefont{Menendez, Gazit,
  and Schwenk}}]{Menendez:2012tm}
\bibinfo{author}{\bibfnamefont{J.}~\bibnamefont{Menendez}},
  \bibinfo{author}{\bibfnamefont{D.}~\bibnamefont{Gazit}}, \bibnamefont{and}
  \bibinfo{author}{\bibfnamefont{A.}~\bibnamefont{Schwenk}},
  \bibinfo{journal}{Phys. Rev.} \textbf{\bibinfo{volume}{D86}},
  \bibinfo{pages}{103511} (\bibinfo{year}{2012}), \eprint{1208.1094}.

\bibitem[{\citenamefont{Klos et~al.}(2013)\citenamefont{Klos, Menendez, Gazit,
  and Schwenk}}]{Klos:2013rwa}
\bibinfo{author}{\bibfnamefont{P.}~\bibnamefont{Klos}},
  \bibinfo{author}{\bibfnamefont{J.}~\bibnamefont{Menendez}},
  \bibinfo{author}{\bibfnamefont{D.}~\bibnamefont{Gazit}}, \bibnamefont{and}
  \bibinfo{author}{\bibfnamefont{A.}~\bibnamefont{Schwenk}},
  \bibinfo{journal}{Phys. Rev.} \textbf{\bibinfo{volume}{D88}},
  \bibinfo{pages}{083516} (\bibinfo{year}{2013}), \bibinfo{note}{[Erratum:
  Phys. Rev.D89,no.2,029901(2014)]}, \eprint{1304.7684}.

\bibitem[{\citenamefont{Baudis et~al.}(2013)\citenamefont{Baudis, Kessler,
  Klos, Lang, Menendez, Reichard, and Schwenk}}]{Baudis:2013bba}
\bibinfo{author}{\bibfnamefont{L.}~\bibnamefont{Baudis}},
  \bibinfo{author}{\bibfnamefont{G.}~\bibnamefont{Kessler}},
  \bibinfo{author}{\bibfnamefont{P.}~\bibnamefont{Klos}},
  \bibinfo{author}{\bibfnamefont{R.~F.} \bibnamefont{Lang}},
  \bibinfo{author}{\bibfnamefont{J.}~\bibnamefont{Menendez}},
  \bibinfo{author}{\bibfnamefont{S.}~\bibnamefont{Reichard}}, \bibnamefont{and}
  \bibinfo{author}{\bibfnamefont{A.}~\bibnamefont{Schwenk}},
  \bibinfo{journal}{Phys. Rev.} \textbf{\bibinfo{volume}{D88}},
  \bibinfo{pages}{115014} (\bibinfo{year}{2013}), \eprint{1309.0825}.

\bibitem[{\citenamefont{Vietze et~al.}(2015)\citenamefont{Vietze, Klos,
  Menendez, Haxton, and Schwenk}}]{Vietze:2014vsa}
\bibinfo{author}{\bibfnamefont{L.}~\bibnamefont{Vietze}},
  \bibinfo{author}{\bibfnamefont{P.}~\bibnamefont{Klos}},
  \bibinfo{author}{\bibfnamefont{J.}~\bibnamefont{Menendez}},
  \bibinfo{author}{\bibfnamefont{W.~C.} \bibnamefont{Haxton}},
  \bibnamefont{and} \bibinfo{author}{\bibfnamefont{A.}~\bibnamefont{Schwenk}},
  \bibinfo{journal}{Phys. Rev.} \textbf{\bibinfo{volume}{D91}},
  \bibinfo{pages}{043520} (\bibinfo{year}{2015}), \eprint{1412.6091}.

\bibitem[{\citenamefont{Goodman et~al.}(2011)\citenamefont{Goodman, Ibe,
  Rajaraman, Shepherd, Tait, and Yu}}]{Goodman:2010qn}
\bibinfo{author}{\bibfnamefont{J.}~\bibnamefont{Goodman}},
  \bibinfo{author}{\bibfnamefont{M.}~\bibnamefont{Ibe}},
  \bibinfo{author}{\bibfnamefont{A.}~\bibnamefont{Rajaraman}},
  \bibinfo{author}{\bibfnamefont{W.}~\bibnamefont{Shepherd}},
  \bibinfo{author}{\bibfnamefont{T.~M.~P.} \bibnamefont{Tait}},
  \bibnamefont{and} \bibinfo{author}{\bibfnamefont{H.-B.} \bibnamefont{Yu}},
  \bibinfo{journal}{Nucl. Phys.} \textbf{\bibinfo{volume}{B844}},
  \bibinfo{pages}{55} (\bibinfo{year}{2011}), \eprint{1009.0008}.

\bibitem[{\citenamefont{Bishara
  et~al.}(2017{\natexlab{c}})\citenamefont{Bishara, Brod, Grinstein, and
  Zupan}}]{Bishara:2017nnn}
\bibinfo{author}{\bibfnamefont{F.}~\bibnamefont{Bishara}},
  \bibinfo{author}{\bibfnamefont{J.}~\bibnamefont{Brod}},
  \bibinfo{author}{\bibfnamefont{B.}~\bibnamefont{Grinstein}},
  \bibnamefont{and} \bibinfo{author}{\bibfnamefont{J.}~\bibnamefont{Zupan}}
  (\bibinfo{year}{2017}{\natexlab{c}}), \eprint{1708.02678}.

\bibitem[{\citenamefont{Lehman and Martin}(2015)}]{Lehman:2015via}
\bibinfo{author}{\bibfnamefont{L.}~\bibnamefont{Lehman}} \bibnamefont{and}
  \bibinfo{author}{\bibfnamefont{A.}~\bibnamefont{Martin}},
  \bibinfo{journal}{Phys. Rev.} \textbf{\bibinfo{volume}{D91}},
  \bibinfo{pages}{105014} (\bibinfo{year}{2015}), \eprint{1503.07537}.

\bibitem[{\citenamefont{Henning et~al.}(2016)\citenamefont{Henning, Lu, Melia,
  and Murayama}}]{Henning:2015daa}
\bibinfo{author}{\bibfnamefont{B.}~\bibnamefont{Henning}},
  \bibinfo{author}{\bibfnamefont{X.}~\bibnamefont{Lu}},
  \bibinfo{author}{\bibfnamefont{T.}~\bibnamefont{Melia}}, \bibnamefont{and}
  \bibinfo{author}{\bibfnamefont{H.}~\bibnamefont{Murayama}},
  \bibinfo{journal}{Commun. Math. Phys.} \textbf{\bibinfo{volume}{347}},
  \bibinfo{pages}{363} (\bibinfo{year}{2016}), \eprint{1507.07240}.

\bibitem[{\citenamefont{Henning et~al.}(2017)\citenamefont{Henning, Lu, Melia,
  and Murayama}}]{Henning:2015alf}
\bibinfo{author}{\bibfnamefont{B.}~\bibnamefont{Henning}},
  \bibinfo{author}{\bibfnamefont{X.}~\bibnamefont{Lu}},
  \bibinfo{author}{\bibfnamefont{T.}~\bibnamefont{Melia}}, \bibnamefont{and}
  \bibinfo{author}{\bibfnamefont{H.}~\bibnamefont{Murayama}},
  \bibinfo{journal}{JHEP} \textbf{\bibinfo{volume}{08}}, \bibinfo{pages}{016}
  (\bibinfo{year}{2017}), \eprint{1512.03433}.

\bibitem[{\citenamefont{Brod et~al.}(2018)\citenamefont{Brod, Grinstein,
  Stamou, and Zupan}}]{Brod:2018ust}
\bibinfo{author}{\bibfnamefont{J.}~\bibnamefont{Brod}},
  \bibinfo{author}{\bibfnamefont{B.}~\bibnamefont{Grinstein}},
  \bibinfo{author}{\bibfnamefont{E.}~\bibnamefont{Stamou}}, \bibnamefont{and}
  \bibinfo{author}{\bibfnamefont{J.}~\bibnamefont{Zupan}},
  \bibinfo{journal}{JHEP} \textbf{\bibinfo{volume}{02}}, \bibinfo{pages}{174}
  (\bibinfo{year}{2018}), \eprint{1801.04240}.

\bibitem[{\citenamefont{Ovanesyan and Vecchi}(2015)}]{Ovanesyan:2014fha}
\bibinfo{author}{\bibfnamefont{G.}~\bibnamefont{Ovanesyan}} \bibnamefont{and}
  \bibinfo{author}{\bibfnamefont{L.}~\bibnamefont{Vecchi}},
  \bibinfo{journal}{JHEP} \textbf{\bibinfo{volume}{07}}, \bibinfo{pages}{128}
  (\bibinfo{year}{2015}), \eprint{1410.0601}.

\bibitem[{\citenamefont{Appelquist et~al.}(2015)}]{Appelquist:2015zfa}
\bibinfo{author}{\bibfnamefont{T.}~\bibnamefont{Appelquist}}
  \bibnamefont{et~al.}, \bibinfo{journal}{Phys. Rev. Lett.}
  \textbf{\bibinfo{volume}{115}}, \bibinfo{pages}{171803}
  (\bibinfo{year}{2015}), \eprint{1503.04205}.

\bibitem[{\citenamefont{Weiner and Yavin}(2012)}]{Weiner:2012cb}
\bibinfo{author}{\bibfnamefont{N.}~\bibnamefont{Weiner}} \bibnamefont{and}
  \bibinfo{author}{\bibfnamefont{I.}~\bibnamefont{Yavin}},
  \bibinfo{journal}{Phys. Rev.} \textbf{\bibinfo{volume}{D86}},
  \bibinfo{pages}{075021} (\bibinfo{year}{2012}), \eprint{1206.2910}.

\bibitem[{\citenamefont{Frandsen et~al.}(2012)\citenamefont{Frandsen, Haisch,
  Kahlhoefer, Mertsch, and Schmidt-Hoberg}}]{Frandsen:2012db}
\bibinfo{author}{\bibfnamefont{M.~T.} \bibnamefont{Frandsen}},
  \bibinfo{author}{\bibfnamefont{U.}~\bibnamefont{Haisch}},
  \bibinfo{author}{\bibfnamefont{F.}~\bibnamefont{Kahlhoefer}},
  \bibinfo{author}{\bibfnamefont{P.}~\bibnamefont{Mertsch}}, \bibnamefont{and}
  \bibinfo{author}{\bibfnamefont{K.}~\bibnamefont{Schmidt-Hoberg}},
  \bibinfo{journal}{JCAP} \textbf{\bibinfo{volume}{1210}}, \bibinfo{pages}{033}
  (\bibinfo{year}{2012}), \eprint{1207.3971}.

\bibitem[{\citenamefont{Buchmuller and Wyler}(1986)}]{Buchmuller:1985jz}
\bibinfo{author}{\bibfnamefont{W.}~\bibnamefont{Buchmuller}} \bibnamefont{and}
  \bibinfo{author}{\bibfnamefont{D.}~\bibnamefont{Wyler}},
  \bibinfo{journal}{Nucl. Phys.} \textbf{\bibinfo{volume}{B268}},
  \bibinfo{pages}{621} (\bibinfo{year}{1986}).

\bibitem[{\citenamefont{Grzadkowski et~al.}(2010)\citenamefont{Grzadkowski,
  Iskrzynski, Misiak, and Rosiek}}]{Grzadkowski:2010es}
\bibinfo{author}{\bibfnamefont{B.}~\bibnamefont{Grzadkowski}},
  \bibinfo{author}{\bibfnamefont{M.}~\bibnamefont{Iskrzynski}},
  \bibinfo{author}{\bibfnamefont{M.}~\bibnamefont{Misiak}}, \bibnamefont{and}
  \bibinfo{author}{\bibfnamefont{J.}~\bibnamefont{Rosiek}},
  \bibinfo{journal}{JHEP} \textbf{\bibinfo{volume}{10}}, \bibinfo{pages}{085}
  (\bibinfo{year}{2010}), \eprint{1008.4884}.

\bibitem[{\citenamefont{Bishara et~al.}(2018)\citenamefont{Bishara, Brod,
  Grinstein, and Zupan}}]{Bishara:2018vix}
\bibinfo{author}{\bibfnamefont{F.}~\bibnamefont{Bishara}},
  \bibinfo{author}{\bibfnamefont{J.}~\bibnamefont{Brod}},
  \bibinfo{author}{\bibfnamefont{B.}~\bibnamefont{Grinstein}},
  \bibnamefont{and} \bibinfo{author}{\bibfnamefont{J.}~\bibnamefont{Zupan}}
  (\bibinfo{year}{2018}), \eprint{1809.03506}.

\bibitem[{\citenamefont{Denner}(1993)}]{Denner:1991kt}
\bibinfo{author}{\bibfnamefont{A.}~\bibnamefont{Denner}},
  \bibinfo{journal}{Fortsch.Phys.} \textbf{\bibinfo{volume}{41}},
  \bibinfo{pages}{307} (\bibinfo{year}{1993}), \eprint{0709.1075}.

\bibitem[{\citenamefont{Fedderke et~al.}(2014)\citenamefont{Fedderke, Chen,
  Kolb, and Wang}}]{Fedderke:2014wda}
\bibinfo{author}{\bibfnamefont{M.~A.} \bibnamefont{Fedderke}},
  \bibinfo{author}{\bibfnamefont{J.-Y.} \bibnamefont{Chen}},
  \bibinfo{author}{\bibfnamefont{E.~W.} \bibnamefont{Kolb}}, \bibnamefont{and}
  \bibinfo{author}{\bibfnamefont{L.-T.} \bibnamefont{Wang}},
  \bibinfo{journal}{JHEP} \textbf{\bibinfo{volume}{08}}, \bibinfo{pages}{122}
  (\bibinfo{year}{2014}), \eprint{1404.2283}.

\bibitem[{\citenamefont{Weiner and Yavin}(2013)}]{Weiner:2012gm}
\bibinfo{author}{\bibfnamefont{N.}~\bibnamefont{Weiner}} \bibnamefont{and}
  \bibinfo{author}{\bibfnamefont{I.}~\bibnamefont{Yavin}},
  \bibinfo{journal}{Phys. Rev.} \textbf{\bibinfo{volume}{D87}},
  \bibinfo{pages}{023523} (\bibinfo{year}{2013}), \eprint{1209.1093}.

\bibitem[{\citenamefont{Gripaios and Sutherland}(2016)}]{Gripaios:2016xuo}
\bibinfo{author}{\bibfnamefont{B.}~\bibnamefont{Gripaios}} \bibnamefont{and}
  \bibinfo{author}{\bibfnamefont{D.}~\bibnamefont{Sutherland}},
  \bibinfo{journal}{JHEP} \textbf{\bibinfo{volume}{08}}, \bibinfo{pages}{103}
  (\bibinfo{year}{2016}), \eprint{1604.07365}.

\bibitem[{\citenamefont{Gootjes-Dreesbach}(2016)}]{AWGD:master-thesis}
\bibinfo{author}{\bibfnamefont{A.}~\bibnamefont{Gootjes-Dreesbach}}, Master's
  thesis, \bibinfo{school}{TU Dortmund} (\bibinfo{year}{2016}),
  \bibinfo{note}{unpublished}.

\bibitem[{\citenamefont{Nishi}(2005)}]{Nishi:2004st}
\bibinfo{author}{\bibfnamefont{C.~C.} \bibnamefont{Nishi}},
  \bibinfo{journal}{Am. J. Phys.} \textbf{\bibinfo{volume}{73}},
  \bibinfo{pages}{1160} (\bibinfo{year}{2005}), \eprint{hep-ph/0412245}.

\bibitem[{\citenamefont{Borodulin et~al.}(2017)\citenamefont{Borodulin,
  Rogalyov, and Slabospitskii}}]{Borodulin:2017pwh}
\bibinfo{author}{\bibfnamefont{V.~I.} \bibnamefont{Borodulin}},
  \bibinfo{author}{\bibfnamefont{R.~N.} \bibnamefont{Rogalyov}},
  \bibnamefont{and} \bibinfo{author}{\bibfnamefont{S.~R.}
  \bibnamefont{Slabospitskii}} (\bibinfo{year}{2017}), \eprint{1702.08246}.

\bibitem[{\citenamefont{Dreiner et~al.}(2010)\citenamefont{Dreiner, Haber, and
  Martin}}]{Dreiner:2008tw}
\bibinfo{author}{\bibfnamefont{H.~K.} \bibnamefont{Dreiner}},
  \bibinfo{author}{\bibfnamefont{H.~E.} \bibnamefont{Haber}}, \bibnamefont{and}
  \bibinfo{author}{\bibfnamefont{S.~P.} \bibnamefont{Martin}},
  \bibinfo{journal}{Phys. Rept.} \textbf{\bibinfo{volume}{494}},
  \bibinfo{pages}{1} (\bibinfo{year}{2010}), \eprint{0812.1594}.

\bibitem[{\citenamefont{Simma}(1994)}]{Simma:1993ky}
\bibinfo{author}{\bibfnamefont{H.}~\bibnamefont{Simma}},
  \bibinfo{journal}{Z.Phys.} \textbf{\bibinfo{volume}{C61}},
  \bibinfo{pages}{67} (\bibinfo{year}{1994}), \eprint{hep-ph/9307274}.

\bibitem[{\citenamefont{Neubert}(1994)}]{Neubert:1993mb}
\bibinfo{author}{\bibfnamefont{M.}~\bibnamefont{Neubert}},
  \bibinfo{journal}{Phys. Rept.} \textbf{\bibinfo{volume}{245}},
  \bibinfo{pages}{259} (\bibinfo{year}{1994}), \eprint{hep-ph/9306320}.

\end{thebibliography}

\end{document}